\documentclass[sigconf]{acmart}

\usepackage{enumitem}
\usepackage{tikz}
\usetikzlibrary{arrows.meta, positioning}
\usepackage{polyglossia}
\setmainlanguage{english}
\setotherlanguage{bengali}
\newfontscript{BengaliOpenTypeOld}{beng}
\newfontscript{BengaliOpenTypeNew}{bng2}
\newfontfamily\bengalifont{NotoSansBengali.ttf}[Script=BengaliOpenTypeNew]

\usepackage{fontspec}

\usepackage[T1]{fontenc}
\AtBeginDocument{%
  }

\copyrightyear{2026}
\acmYear{2026}
\setcopyright{cc}
\setcctype{by}
\acmConference[COMPASS '26]{ACM SIGCAS/SIGCHI Conference on Computing and Sustainable Societies}{July 27--31, 2026}{Virtual Event, USA}
\acmBooktitle{ACM SIGCAS/SIGCHI Conference on Computing and Sustainable Societies (COMPASS '26), July 27--31, 2026, Virtual Event, USA}
\acmDOI{10.1145/3811242.3819096}
\acmISBN{979-8-4007-2702-3/2026/07}

\begin{document}

\title[Mod-Guide: An LLM-based Content Moderation Feedback System]{\textit{Mod-Guide}: An LLM-based Content Moderation Feedback System to Address Insensitive Speech toward Indigenous Ethnic and Religious Minority Communities}

\author{Dipto Das}
\affiliation{%
  \department{Department of Computer Science}
  \institution{University of Toronto}
  \city{Toronto}
  \state{Ontario}
  \country{Canada}
}
\email{dipto.das@utoronto.ca}

\author{Achhiya Sultana}
\affiliation{%
  \institution{Independent University Bangladesh}
  \city{Dhaka}
  \country{Bangladesh}
}
\email{achhiyasets@iub.edu.bd}

\author{Ankit Singh Chauhan}
\affiliation{%
  \institution{Indiana University Indianapolis}
  \city{Indianapolis}
  \state{Indiana}
  \country{United States}
}
\email{ankichau@iu.edu}

\author{Saadia Binte Alam}
\affiliation{%
  \institution{Independent University Bangladesh}
  \city{Dhaka}
  \country{Bangladesh}
}
\email{saadiabinte@iub.edu.bd}

\author{Mohammad Shidujaman}
\affiliation{%
  \institution{Independent University Bangladesh}
  \city{Dhaka}
  \country{Bangladesh}
}
\email{shidujaman@iub.edu.bd}

\author{Shion Guha}
\affiliation{%
  \department{Faculty of Information}
  \institution{University of Toronto}
  \city{Toronto}
  \state{Ontario}
  \country{Canada}}
\email{shion.guha@utoronto.ca}

\author{Sunandan Chakraborty}
\affiliation{%
  \institution{Indiana University Indianapolis}
  \city{Indianapolis}
  \state{Indiana}
  \country{United States}
}
\email{sunchak@iu.edu}

\author{Syed Ishtiaque Ahmed}
\affiliation{%
  \department{Department of Computer Science}
  \institution{University of Toronto}
  \city{Toronto}
  \state{Ontario}
  \country{Canada}
}
\email{ishtiaque@cs.toronto.edu}

\renewcommand{\shortauthors}{Das, Sultana, Chauhan, Alam, Shidujaman, Guha, Chakraborty, and Ahmed}

\begin{abstract}
  Language operates as a mechanism of both marginalization and resistance, especially for minority communities navigating insensitive and harmful speech online. As content moderation increasingly depends on large language models (LLMs), concerns arise about whether these systems can recognize culturally insensitive speech--language that disregards or marginalizes the cultural and religious perspectives of historically underrepresented communities, often through implicit erasure, misrepresentation, or normative framing, rather than overt hostility. Focusing on Bangladesh's Hindu and Chakma communities -- the country's largest religious and Indigenous ethnic minorities, respectively -- this paper investigates the epistemic limits of LLM-based moderation systems and explores methods for incorporating minority perspectives. We co-created a culturally grounded corpus of insensitive speech with community members and integrated their narratives into moderation pipelines using retrieval augmented generation (RAG). Our tool, Mod-Guide, improves LLM sensitivity to minority viewpoints by leveraging contextual cues derived from lived experience. Through mixed-method evaluations involving both minority and majority participants, we demonstrate that RAG-enhanced moderation responses are more contextually accurate and perceived differently across ethnic lines. This work advances research in human-computer interaction, AI ethics, and social computing by foregrounding restorative justice and hermeneutical inclusion in the design of content moderation systems.
\end{abstract}

\begin{CCSXML}
<ccs2012>
   <concept>
       <concept_id>10003120.10003121.10003129</concept_id>
       <concept_desc>Human-centered computing~Interactive systems and tools</concept_desc>
       <concept_significance>500</concept_significance>
       </concept>
   <concept>
       <concept_id>10003120.10003130.10011762</concept_id>
       <concept_desc>Human-centered computing~Empirical studies in collaborative and social computing</concept_desc>
       <concept_significance>500</concept_significance>
       </concept>
   <concept>
       <concept_id>10010405.10010497</concept_id>
       <concept_desc>Applied computing~Document management and text processing</concept_desc>
       <concept_significance>300</concept_significance>
       </concept>
   <concept>
       <concept_id>10003456.10010927.10003611</concept_id>
       <concept_desc>Social and professional topics~Race and ethnicity</concept_desc>
       <concept_significance>500</concept_significance>
       </concept>
   <concept>
       <concept_id>10003456.10010927.10003612</concept_id>
       <concept_desc>Social and professional topics~Religious orientation</concept_desc>
       <concept_significance>500</concept_significance>
       </concept>
   <concept>
       <concept_id>10003456.10010927.10003619</concept_id>
       <concept_desc>Social and professional topics~Cultural characteristics</concept_desc>
       <concept_significance>500</concept_significance>
       </concept>
 </ccs2012>
\end{CCSXML}

\ccsdesc[500]{Social and professional topics~Race and ethnicity}
\ccsdesc[500]{Social and professional topics~Religious orientation}
\ccsdesc[500]{Social and professional topics~Cultural characteristics}
\ccsdesc[300]{Human-centered computing~Interactive systems and tools}
\ccsdesc[300]{Human-centered computing~Empirical studies in collaborative and social computing}
\ccsdesc[100]{Applied computing~Document management and text processing}

\keywords{Minority, LLM, RAG, content moderation, ethics}


\maketitle

\section{Introduction}
Language is more than a means of communication and is a form of power~\cite{singh2025power}. It shapes social hierarchies, legitimizes authority, and enables the marginalization--a process through which individuals and groups are pushed to the periphery of society based on attributes like race, gender, ethnicity, religion, caste, nationality, language, sexual orientation, etc.~\cite{erete2018intersectional}. Linguistic marginalization and injuries manifest in online communities through hate speech, bullying, political incitement, and other forms of insensitive speech. In the context of this study, we define insensitive speech as the linguistic acts that--while not overtly hateful or profane--disregards, misrepresents, trivializes, or marginalizes the cultural, religious, or epistemic values of historically underrepresented communities. Unlike hate speech, which is often explicit in its hostility or incitement, insensitive speech manifests through dismissive framings, culturally uninformed generalizations, or normative assumptions rooted in majority worldviews. Its harm arises not only from intent or content, but from its failure to recognize and respect the situated meanings, lived experiences, and interpretive frameworks of minority groups.

Most platforms respond to harmful content by enforcing moderation policies through a combination of human moderators and algorithmic systems~\cite{jiang2023trade, molina2022ai}. Recent advances in large language models (LLMs) have enabled more scalable moderation~\cite{kolla2024llm, zeng2024shieldgemma}, but these models are predominantly shaped by and reinforce majority perspectives~\cite{li2024your}. Given the epistemic underrepresentation of the religious and Indigenous ethnic minorities, whose perspectives and experiences with insensitive or harmful speech might significantly differ from those of the majority in those LLM-based content moderation systems, it would likely reinforce the societal barrier between the majority and minority groups in the case of understanding each other's perspectives.

We focus on Bangladesh, where the Hindu and Chakma communities are the largest religious and Indigenous ethnic minorities, respectively~\cite{bsb2022preliminary}. Motivated by concepts of hermeneutical injustice~\cite{fricker2007epistemic} and the divide between majority and minority consciousness~\cite{du2015souls}, we collaborated with members from those communities to curate a corpus of culturally insensitive statements. Participants described why specific speech acts were hurtful and problematic, grounding their explanations in religious texts, oral histories, cultural practices and rituals, lived experiences, and documents from rights organizations. These insights reflect interpretive resources that are typically excluded from LLM training data. To operationalize these perspectives, we introduce \textbf{\textit{Mod-Guide}}, an LLM-based moderation feedback tool that uses retrieval-augmented generation (RAG) to ground moderation responses in this community-sourced corpus. While RAG has shown strong performance across a range of NLP tasks~\cite{lewis2020retrieval}, the significance of our work lies in grounding RAG in epistemically marginalized perspectives and evaluating its implications for culturally sensitive moderation. We evaluate Mod-Guide using a mixed-method study with participants from the majority and minority communities, comparing its outputs to responses from the off-the-shelf GPT-4 model. Our analysis shows that grounding LLM responses in minority perspectives through RAG significantly affects how harmful speech is interpreted and moderated. We also find that the perceived usefulness of these moderation outputs varies by ethnicity but not by religion. This work makes two key contributions that are widely recognized in HCI scholarship~\cite{wobbrock2012seven}:

\begin{itemize}[leftmargin=*]
    \item \textbf{Dataset contribution}: a curated and annotated corpus of culturally insensitive speech from minority perspectives.
    \item \textbf{Artifact contribution}: the design and evaluation of Mod-Guide, a feedback system that integrates these perspectives into the workflow of LLM-based moderation.
\end{itemize}

This research contributes to ongoing discourse in human-computer interaction, AI ethics, and social computing by centering epistemically marginalized communities in data curation and system design. It demonstrates how LLMs can be made more sensitive to pluralistic norms through community participation and socio-technical design. The following sections detail the sociolinguistic framing of marginalization, the construction of our dataset, the design of the LLM-RAG pipeline, and the empirical evaluation. We conclude by reflecting on the challenges of scale, the normativity of dataset curation, and the implications for design toward community-centered justice and fair content moderation systems.
\section{Literature Review}\label{sec:literature_review}
This section situates our work at the intersection of linguistic marginalization, epistemic injustice, and automated content moderation. First, we discuss linguistic marginalization and conceptualize \emph{insensitive speech} as a form of harm shaped by cultural and historical context. Next, we examine epistemic barriers between majority and minority communities through the lenses of Du Bois' notion of the \emph{veil} and Fricker's \emph{hermeneutical injustice}. Finally, we review research on automated content moderation and large language models, highlighting limitations in addressing culturally contextual harms and motivating our community-grounded RAG-based approach.

\subsection{Linguistic Marginalization as Insensitive Speech}
Language plays a crucial role in shaping social hierarchies and power dynamics. It establishes normative and non-normative identities~\cite{butler2021excitable}. As such, people are marginalized through language, often in the form of bullying, hate speech, and threats. Similarly, religious and ethnic minorities are also vulnerable to linguistic injuries. Such injury arises not only from offensive speech targeting certain religions and ethnicities but also from the mode or ways those identities are positioned as dismissed and devalued~\cite{butler2021excitable}. In this paper, we focus on linguistic injuries and vulnerabilities, where exact words may not be explicitly offensive (e.g., name-calling), yet their conventional bearing--how words derive power from historical and social conventions—can come across as disregarding or diminishing the experiences, identities, practices, and contexts of religious and ethnic minorities, which we dub as insensitive speech.

% For example, the experience of people who identify as lesbian, gay, bisexual, transgender, and queer (LGBTQ) disclosing their sexual and/or gender identity can be stressful and even traumatic, as their identity faces verbal and physical discrimination across social settings, including home, school, and broader social environments~\cite{dym2019coming}.

To study the linguistic marginalization of religious and ethnic minorities in Bangladesh, we need to understand their sociopolitical contexts. Religious minorities in Bangladesh, particularly Hindus, have long faced marginalization characterized by both historical and ongoing violence~\cite{rifat2024politics}. The large-scale communal riots and the disproportionate targeting of Hindus during the Liberation War illustrate this pattern~\cite{sarkar2017calcutta, anam2013pakistan}. In recent decades, assaults on Hindu communities have increased, often fueled by social media rumors of religious insults against the majority~\cite{ganguly2021bangladesh, roy2023sociological, amnesty2021bangladesh}, such as the violence during the 2021 Durga Puja~\cite{hasan2021minorities}. Furthermore, political instability worsens the persecution, leading to targeted attacks on Hindus~\cite{ittefaq2014attacks, prothomalo2024minority}, Christians~\cite{minority2018christians}, and atheists~\cite{theguardianAmerican2015Atheist, shackle2018atheist}. Similarly, the Indigenous ethnic minorities in Bangladesh (known as \textit{Adivasi}) face marginalization due to their ethnic and cultural differences from the majority Bengali population. These communities, particularly in the Chittagong Hill Tracts, have experienced displacement, settlement, encroachment on their ancestral lands, ethnocide, and violence due to the region’s militarization since before the country's independence~\cite{chakma2010post, hill2022muscular}. Despite a peace accord in 1997, they continue to struggle for autonomy and basic recognition of indigeneity to this day~\cite{chakma2008assessing, dailystar2025removal}.

Recent scholarships in social computing and ICT for development have looked into how these sociopolitical experiences of religious and ethnic minorities in Bangladesh manifest as everyday linguistic marginalization in online communities in their interaction with other users and content moderation. For example,~\cite{rifat2024politics} explained how social psychology shapes the participation of religious minorities online, who, due to a fear of isolation, fall into a spiral of silence, negotiate through the future uncertainties and present impression of fear, and accommodate their communication with religious majority communities. Among the Indigenous communities in Bangladesh, many share religious minority identities, such as Chakma, Santhals, and Garo, who follow Buddhism, Hinduism, and Christianity, respectively~\cite{theworldinusIndigenousPeoples}. Users from these communities have markedly different experiences with hate speech on online platforms compared to their peers from the majority community. The lack of urgency in addressing their experience with explicitly profane speech creates a clear disparity concerning membership, rights, and participation as users of online platforms~\cite{sultana2024civics}. Taking that into account, efforts to address insensitive speech with conventional bearing are more likely to be influenced by majoritarianism and, hence, require additional contextual content moderation and depend on increased awareness among majority religious and ethnic groups, such as the Bengali Muslims in Bangladesh.

\subsection{Epistemic Barriers among Majority and Minority}
Marginalization of minorities often stems from entrenched tribal stigma surrounding attributes like ethnicity, religion, language, and cultural practices~\cite{goffman2009stigma}. For example, in many contexts, misunderstandings of minority religions' practices and beliefs lead to unsubstantiated fear (e.g., Islamophobia~\cite{allen2016islamophobia}), misrepresentation (e.g., depicting non-Abrahamic faiths as satanic or pagan~\cite{sugirtharajah2004imagining}), or exclusion. Similarly, immigrants who speak different languages often face suspicion or hostility, as their speech is perceived as secretive or exclusionary, reinforcing their marginalization in the form of xenophobia~\cite{lee2019america}. Scholars argue that such stigma and marginalization are not the victims' attributes but a feature of the society that imposes it. Through various social processes, minorities' symbols, beliefs, practices, and physical conditions are made non-normative in society and are devalued or discredited to such an extent that they adopt different coping mechanisms~\cite{goffman2009stigma}, such as hiding their identities, avoiding sharing their experiences or withdrawing from social interactions out of fear of isolation and the desire to conform to norms in both online and offline settings~\cite{rifat2024politics}.

In this paper, we seek to understand the experiences of religious and ethnic minorities being marginalized, ridiculed, and misunderstood in the Bangladeshi social media sphere by combining W.E.B. Du Bois' concept of ``the veil''~\cite{du2015souls} and Miranda Fricker's notion of hermeneutical injustice~\cite{fricker2007epistemic}. These theoretical angles provide complementary lenses for understanding and addressing the underlying processes that lead to the minorities' marginalization. Du Bois' conceptualization of the ``veil'' highlights how racial minorities in the United States experience an imposed separation that distorts their self-perception and hinders mutual comprehension across racial divides~\cite{du2015souls}. Recent work~\cite{rifat2024politics} in the context of Bangladesh has highlighted how the religious minority communities feel a comparable divide between themselves and the religious majority, particularly in how their identities and practices are misinterpreted, leading to alienation and marginalization. That metaphorical veil between the majority and minority groups in terms of ethnicity or religion functions as an epistemic barrier, preventing adequate and effective intergroup understanding.

Drawing from Fricker's work~\cite{fricker2007epistemic}, this epistemic difference could be dubbed hermeneutical injustice, where minority groups struggle to make sense of their experiences due to the lack of necessary conceptual resources within normative epistemic frameworks shaped by religious and ethnic majorities' beliefs and practices. For example, theological interpretations (e.g., the role of idols in worship for Hindu minorities) and dietary practices of the ethnic minority communities (e.g., consumption of pork, frog, and alcohol) are considered wrong from the perspective of the majority Bengali Muslims' standpoint~\cite{rifat2024politics, sultana2024civics, sultana2022toleration}. When members of the majority community talk about those beliefs and practices, the minority groups might deem such comments as stereotypical, condescending, insulting, and overall insensitive, which reinforces division and further marginalizes minorities online.

Divisions between majority and minority groups are sustained by institutionalized ignorance and a lack of empathy~\cite{du2015souls}, while dominant social norms and unconscious biases perpetuate injustice against marginalized communities~\cite{fricker2007epistemic}. In online communities where religious and ethnic minorities encounter insensitive speech, different moderation and feedback mechanisms could be implemented with careful attention to the epistemologies of these groups. More broadly, dismantling these barriers demands inclusive epistemic practices—encompassing knowledge production, recognition, and validation—to value minorities' perspectives and foster interfaith communication and mutual understanding. These practices would ultimately address the power asymmetries experienced by religious and ethnic minorities online by shaping the design and governance of sociotechnical systems like online platforms.

\subsection{Language Models in Moderating Insensitive Speech}
With the global adoption of online platforms and the diverse communities they host, moderating harmful and insensitive speech has become a complex sociotechnical challenge. Existing scholarship has shown that perceptions of what constitutes harmful content and its severity vary significantly across cultural and social contexts~\cite{jiang2021understanding, scheuerman2021framework}. While platforms' ``institutional ethics''~\cite{scheuerman2021framework} do not want to implement the perspectives of users who think anything that does not pertain to a particular religious belief should be removed, they rarely make an active effort in addressing the hermeneutical injustice~\cite{jiang2021understanding}, i.e., the structural exclusion of minority perspectives in defining what counts as harmful. As online communities grow, platforms must negotiate competing moderation values (e.g., community identity), philosophies (e.g., nurturing vs. punishing), and implementation styles (e.g., human vs. algorithmic moderation)~\cite{jiang2023trade, das2021jol}.

Particularly focused on moderation philosophy, Seering et al.~\cite{seering2022metaphors} examined how moderation can be conceptualized through different metaphors, such as mentoring, law enforcement, and custodianship. These metaphors shape how platforms and moderators perceive their roles, influencing decisions and ethics about intervention, the balance between users' autonomy and governance, and the prioritization of different cultures and values. As the platforms adopt algorithmic moderation for the sake of efficiency, these societal complexities are often pawned off to algorithmic systems~\cite{jiang2023trade}. Language technologies have become central to automated content moderation systems\cite{sun2022design, vaidya2021conceptualizing}. In terms of complexity and sophistication, these systems range from simple keyword filters\cite{jhaver2019human, jhaver2022designing}, to task-specific models for sentiment analysis and hate speech detection~\cite{das2024colonial, mozafari2020hate}, to foundational large language models (LLMs) deployed at scale~\cite{kolla2024llm, zeng2024shieldgemma, inan2024spring}. While multilingual LLMs have shown promising results in detecting explicit hate speech, fake news, and discriminatory language~\cite{plaza2023respectful, koka2024combining, orlandi2021international}, they often struggle with more subtle forms of disinformation and culturally coded insensitivity.

However, LLMs reflect and reinforce dominant cultural norms, which can lead to representational harms, particularly for non-Western communities~\cite{ghosh2024generative, brown2024qualitative}. Prior research has shown that these models exhibit demographic (e.g., race, gender, nationality, religion, caste)~\cite{ghosh2023chatgpt, ghosh2023person, hamidieh2024identifying, ghosh2024interpretations, das2024colonial}, socioeconomic~\cite{arzaghi2024understanding}, and political biases~\cite{agiza2024politune}, raising concerns about how automated moderation disproportionately impacts marginalized communities. Hence, recent works have attempted to reconceptualize moderation by embedding safety paradigms directly into LLM pipelines~\cite{inan2024spring, arzberger2024nothing}, wherein they have examined how data selection and fine-tuning impacted LLMs' economic and political biases~\cite{agiza2024politune}, how model responses vary with culturally sensitive prompts~\cite{mukherjee2024arsenic}, and found that persona-based prompting can improve alignment with specific moderation goals~\cite{kwok2024evaluating}. Studies highlighted how crowd-sourced data annotation is subject to limited annotator expertise~\cite{kumar2024socio}, dismissal of religious faiths~\cite{rifat2024data}, minorities' underrepresentation~\cite{thorne2022data, song2025participant}, and disproportionate association of toxicity with minorities~\cite{wiegand2021implicitly}. Retrieval augmented generation (RAG)--a method to enhance language model outputs by retrieving relevant external documents while generating responses~\cite{lewis2020retrieval}, can be an effective technique to address the concerns of LLM biases affecting content moderation~\cite{leitner2025characterizing, tsirmpas2025scalable}. However, there is a dearth of literature that has examined its effectiveness in moderating content around minority identity, especially in non-English languages and the Global South contexts.

Our work advances research at the intersection of content moderation, LLMs, and low-resource language communities in two key ways. First, we address the dataset challenge by constructing a culturally grounded corpus of insensitive speech in Bengali, annotated and contextualized by members of underrepresented religious and ethnic minority communities in Bangladesh. Rather than relying on crowd-sourced or majority-labels that often obscure minority perspectives, our approach centers the lived experiences, interpretive frameworks, and rationales of those most affected by marginalization. Second, we build on insights from prior literature that persona-based prompting may help align LLM outputs with specific moderation philosophies~\cite{seering2022metaphors, kwok2024evaluating} and RAG enhances factual accuracy and contextual grounding~\cite{izacard2020leveraging, shi2023replug}. We implemented this insight in our content moderation feedback system, Mod-Guide, in which we prompt an LLM to adopt various moderation roles and ground its responses in the minority community-sourced corpus using RAG. We evaluated which configurations--combinations of prompts and the presence/absence of RAG--produce more contextually sensitive, factually accurate, and epistemically inclusive feedback.
\section{Methods}
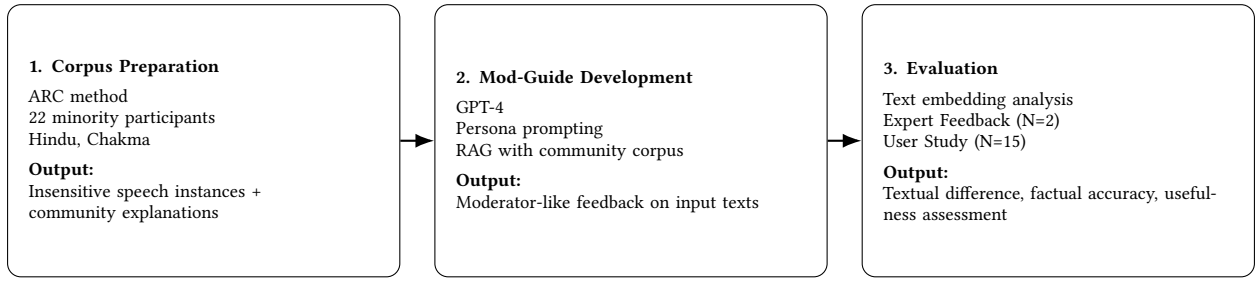
\begin{figure*}[!ht]
\centering
\begin{tikzpicture}[
    font=\footnotesize,
    box/.style={
        draw,
        rounded corners=5pt,
        align=left,
        text width=0.26\linewidth,
        minimum height=3.6cm,
        inner sep=8pt
    },
    arrow/.style={-{Latex[length=2.5mm]}, thick}
]

\node[box] (corpus) {
\textbf{1. Corpus Preparation}\vspace{0.5em}

ARC method\\
22 minority participants\\
Hindu, Chakma

\vspace{0.5em}
\textbf{Output:}\\
Insensitive speech instances +\\
community explanations
};

\node[box, right=0.45cm of corpus] (system) {
\textbf{2. Mod-Guide Development}\vspace{0.5em}

GPT-4\\
Persona prompting\\
RAG with community corpus

\vspace{0.5em}
\textbf{Output:}\\
Moderator-like feedback on input texts
};

\node[box, right=0.45cm of system] (eval) {
\textbf{3. Evaluation}\vspace{0.5em}

Text embedding analysis\\
Expert Feedback (N=2)\\
User Study (N=15)

\vspace{0.5em}
\textbf{Output:}\\
Textual difference, factual accuracy, usefulness assessment
};

\draw[arrow] (corpus.east) -- (system.west);
\draw[arrow] (system.east) -- (eval.west);

\end{tikzpicture}
\caption{Methods overview.}
\label{fig:methods_overview}
\end{figure*}

This paper is part of a broader study to understand minority communities' experiences with content moderation in online communities and develop tools to make those spaces more inclusive and accessible for these communities~\cite{rifat2024data, sultana2024civics, das2021jol}. Here, we build on our findings and community relationships fostered during the earlier phases of our research. Our study proceeded in three stages (see Figure~\ref{fig:methods_overview}): (1) corpus preparation, (2) development of the Mod-Guide moderation system, and (3) evaluation of moderation feedback.

\subsection{Overview}
First, we collaborated with 22 members from two minority communities in Bangladesh--Hindu and Chakma--using the asynchronous research community method to construct a corpus of culturally insensitive speech containing 132 instances and accompanying explanations grounded in community perspectives. Second, we integrated this corpus into a moderation pipeline that combines persona-based prompting GPT-4 with retrieval-augmented generation (RAG) to ground moderation feedback in community-authored explanations. Third, we evaluated the system through a mixed-method approach consisting of (a) quantitative analysis of generated responses using text embeddings, (b) assessment of factual accuracy of generated texts by 2 experts, and (c) a user study examining the perceived usefulness of moderation feedback with 15 participants from majority and minority communities. The following sections offer further details about each of these stages. 

\subsection{Author Positionality}
Prior research has highlighted how the researchers' identities may reflexively address inevitable tensions and bring affinities into perspective in studying marginalized communities~\cite{schlesinger2017intersectional, liang2021embracing}. Therefore, we consider it essential to situate this work on marginalized minority communities in the Global South in relation to the researchers' positionality. Among all authors (2 women and 6 men), five were born and raised in Bangladesh, while the other three were from India. Except for one author (who is from a North Indian ethnic background), all authors belong to the Bengali ethnolinguistic group. Three authors identify as Bengali Hindus (the lead author from an underprivileged caste in Bangladesh, the rest from a dominant caste in West Bengal, India), and four authors were born in Muslim communities. In addition to their varied sociocultural perspectives, all authors' backgrounds in computer science, with different authors' prior research with marginalized communities, text mining, and data science, have informed and guided the motivation and execution of this study.
\section{Corpus Preparation to Understand Minority Hermeneutics}\label{sec:corpus}
We collected the corpus through the Asynchronous Remote Community (ARC) method~\cite{macleod2016asynchronous}. Prior research has used this method to engage with participants when in-person communication is difficult to arrange due to population distribution~\cite{macleod2017grateful}, stigma~\cite{maestre2018defining}, or fear of isolation~\cite{walker2020more}. Over a month, we nudged discussions from those groups weekly to sustain engagement while allowing flexibility for participants to share instances of social media posts they found culturally insensitive, about which they have found that the religious and ethnic majority communities have different perceptions. In doing so, our corpus prioritizes minority hermeneutics--interpretation of their practices, experiences, values, and beliefs from their own perspectives, over being shaped or constrained by majoritarian normative societal views.

\subsection{ARC Participants}
In this paper, we focus on the religious minority Hindu community and the Indigenous ethnic minority Chakma community in Bangladesh. We recruited participants aged 18 years and older by sharing the recruitment materials and additional information with our personal networks, through Facebook advertisements, and by reaching out to participants from our previous studies involving these communities. We also contacted the administrators and moderators of local Facebook groups dedicated to these minority communities, asking their permission to post the call for participation in those groups. We asked the respondents to the study's advertisements to self-identify key characteristics such as gender, caste, age, and their places of upbringing and current residence, which prior studies found to have differing experiences within the Hindu and Chakma communities~\cite{sultana2024civics, rifat2024politics}. Our ARCs with these participants included 11 from the religious minority Hindu community (7 male and 4 female) and 11 from the Indigenous ethnic minority Chakma community (2 male, 4 female, and 5 did not respond to the question asking their gender). Both ARCs had more members, but those who did not post at least once in the groups were excluded from the reported counts. While our religious minority participants came from various parts of the country, most of our ethnic minority participants were from the Chittagong Hill Tracts (CHT) region, where most Indigenous ethnic minority communities live. Most of our Hindu participants were from underprivileged scheduled castes (\textit{tafsili jati})~\cite{sen2018decline}. In addition to reflecting the general demographic pattern of Hindu communities in Bangladesh, the higher representation of participants from underprivileged castes also resists the Brahminical and casteist interpretations of Hindu beliefs and practices in our corpus.

% Participants ranged from X to X years old (average=X, SD=X). For participant details, see Tables~\ref{tab:religious_minorities_demographies} and~\ref{tab:ethnic_minorities_demographies} below. 

% \begin{table}[!ht]
%     \centering
%     \caption{Demographic information about the participants from the religious minority Hindu communities.}
%     \begin{tabular}{p{1cm}p{1.5cm}p{1cm}p{3cm}p{4cm}}
%     \hline
%         \textbf{ID} & \textbf{Gender} & \textbf{Age} & \textbf{Caste} & \textbf{Location} \\\hline
%          & \\\hline
%     \end{tabular}
%     \label{tab:religious_minorities_demographies}
% \end{table}

% \begin{table}[!ht]
%     \centering
%     \caption{Demographic information about the participants from the ethnic minority Chakma communities.}
%     \begin{tabular}{p{1cm}p{1.5cm}p{1cm}p{7cm}}
%     \hline
%         \textbf{ID} & \textbf{Gender} & \textbf{Age} & \textbf{Location} \\\hline
%          & \\\hline
%     \end{tabular}
%     \label{tab:ethnic_minorities_demographies}
% \end{table}

\subsection{Procedure}
Similar to previous ARC studies~\cite{macleod2016asynchronous, maestre2018defining, walker2020more}, based on our participants' preferences, we used a secret Facebook group and a secret WhatsApp group, respectively, to interact with the former and the latter minority groups. Hosting the ARCs on these online platforms minimized the need to familiarize participants with a new system~\cite{macleod2017grateful, heywood2024reaching}. All participants had existing Facebook and WhatsApp accounts that they used to participate in the study, thus maintaining platform-related risks similar to those participants regularly assume while using these communication channels. After completing our informed consent procedure and orienting them with a code of conduct, we invited them to join the groups. From 25/10/2024 to 23/11/2024, we maintained engagement through weekly elicitation while allowing for flexibility.

However, a few participants either did not actively engage or ceased participating after the first couple of weeks in those Facebook and WhatsApp groups, which is a pattern of attrition and participation consistent with previous ARC studies~\cite{prabhakar2017investigating, walker2020more}. The other participants responded to our prompts by sharing examples of textual posts, comments, images, and videos they perceived as insensitive to their religious and Indigenous ethnic identities, cultures, rituals, and practices. We specifically sought instances that were often dismissed as non-problematic by the religious and ethnic majority communities, as the participants experienced through interacting with friends and acquaintances in those communities or having their reports of such content overlooked by content moderation systems on online platforms. We also asked the participants to explain why they found the contents insensitive, referencing sources such as the scriptures of the religious minority communities, national and international resolutions regarding the rights and concerns of the Indigenous ethnic minorities, and their lived experiences and understanding of their respective communities. The participants also engaged with and built upon each others' responses. The first two authors monitored the groups to ensure compliance with the code of conduct and asked follow-up questions to nudge the participants to share additional details.

Our participants participated in the ARCs using both Bengali and English. To streamline the corpus, we translated all written communication into Bengali, the national language of Bangladesh. Participants occasionally shared screenshots of social media posts they considered insensitive. We utilized optical character recognition (OCR) to convert those images into Bengali text. Similarly, any videos shared by participants as examples were transcribed into Bengali text. For web URLs provided by participants as instances of insensitive speech, we transcribed the content into Bengali. We anonymized all these contents before OCR and transcription. For common examples of insensitive speech, some participants shared links to online repositories containing writings and references explaining why such remarks are inappropriate. We scraped the web pages in those cases, excluding non-textual content (e.g., HTML tags, URLs). While allowing the participants to share screenshots, URLs, and videos made it easier for them to share the examples of insensitive speech they encountered, using OCR, transcription, and cleaning non-natural language components enabled textual standardization, allowing us to convert image-based content into analyzable text for inclusion in both the RAG corpus and the moderation evaluation pipeline and allowed us to preserve contextually rich, vernacular examples that participants considered important.

We gathered 53 instances of insensitive speech directed at the religious minority Hindu community and 79 instances targeting the indigenous ethnic minority Chakma community, organizing them into two separate spreadsheets. Each spreadsheet contains two columns: one listing examples of insensitive speech and the other explaining their inappropriateness. Let's consider the following example text \textit{(later referred as Insensitive Speech Example-1)} that Hindu participants in our ARCs found to be culturally insensitive.

\begin{quote}
    \textbengali{কিছু মানবতার ফেলিওয়ালাদের দেখতেছি, মূর্তি পাহারা দিতে মন্দিরে যাচ্ছে। মূর্তি পাহারা দেওয়ার জন্য ঈমান আনিও নাই, মূর্তি পাহারার পক্ষে আমি নাই। ভাঙ্গা লাগলে ডাক দিয়েন} (I have been seeing some vendors of humanism who are going to temples to guard the idols. I did not bring \textit{imaan} (faith in Islam) for guarding the idols, [and] I am not in favor of guarding the idols. Call [me] if those [idols] need to be broken.)
\end{quote}

Since this example text was collected from the post of a user belonging to the religious majority, it reflects their cultural value and belief: the prohibition on idol worship in Islam. In contrast, in the Hindu faith, idols are viewed as a medium for worship. Consequently, a few of our participants pointed out the aforementioned text that was recently well-circulating in the Bangladeshi social media sphere as insensitive speech. They also explained why they consider it culturally insensitive from different angles. For example, while some participants explained the relevance of idols in Hindu rituals based on references from Hindu scriptures, some others presented arguments informed by their observations of social practices in different religions. For example, an ARC participant shared the following explanations for why the above text was insensitive based on different schools of thought within Hinduism:

\begin{quote}
    There are many formless-theist communities in the world who do not believe in incarnations and do not require any tangible deity or symbol for worship or spiritual practice. Again, some who accept formless-theism still acknowledge the necessity of symbols (such as Om, the Dharma Wheel, or the Star of David) in certain contexts. While they do not accept an external image/idol of God, they still mentally envision some form or symbol within their hearts. On this matter, Swami Vivekananda once said: ``Two types of people do not require forms or idols--those who have no concern for religion at all, and the enlightened beings who have transcended all such states. We exist somewhere in between these two conditions. Internally and externally, we need some form of an idol or image."
\end{quote}

We emphasize that our work does not seek to evaluate different theological beliefs and practices. Rather, we aim to highlight how various cultural, religious, and social values influence people's perceptions of content sensitivity and the roles they expect moderators to fulfill. Hence, we will use this corpus of speech the minority communities viewed as culturally insensitive and the rationales behind such perceptions to inform LLM-based automated content moderation.
\section{\textit{Mod-Guide}: Persona-based LLM Prompting and RAG Pipeline for Moderation Feedback}\label{sec:llm}
This paper investigates the effectiveness of large language models (LLMs) in moderating insensitive speech directed at religious and ethnic minority communities in Bangladesh, which is often based on stereotypes and deepens the cultural divide between the majority and minority communities in the country. Drawing on Du Bois~\cite{du2015souls}, we refer to that as the veil. We examined OpenAI's GPT-4 in particular. Additionally, we explore retrieval-augmented generation (RAG)~\cite{lewis2020retrieval} based on community insights, with content moderation in mind. We chose RAG over other approaches, such as few-shot prompting or fine-tuning, to ensure interpretability, adaptability, and alignment with community perspectives. RAG allows generated texts to be directly grounded in retrievable, community-authored explanations, preserving traceability and cultural nuance~\cite{chen2024benchmarking}. Unlike fine-tuning, which embeds knowledge irreversibly into model weights, RAG supports modular updates as community insights evolve. This approach preserves traceability, allows the corpus to evolve as communities contribute additional insights, and supports modular updates as new examples are collected. Although RAG introduces computational overhead compared to simple prompting, it offers an interpretable mechanism for integrating minority hermeneutics into moderation feedback, which aligns with the exploratory and design-oriented objectives of this work. We prompted these systems in different ways to shape their responses by defining their role as content moderators (e.g., nurturing, governing) and persona (e.g., teacher, judge) to build a moderation feedback system called ``Mod-Guide."

In this section, we will first discuss the different prompts, then explain the LLM and RAG pipeline, and finally discuss the interface with examples.

\subsection{Prompting}\label{sec:persona_prompts}
We used five prompts in the pipeline. In all these cases, the objective was the same: to understand whether a piece of text would potentially be perceived as hate speech, hurtful, or culturally sensitive. Those prompts featured different roles of moderators of online interaction. Drawing on Seering et al.~\cite{seering2022metaphors}, we considered five different roles content moderators play in moderating online communities, such as (a) nurturing and supporting, (b) overseeing and facilitating, (c) governing and regulating, (d) managing, and (e) fighting for communities. Compared to other metaphoric conceptualizations of moderation~\cite{zhang2023cleaning, matias2019civic}, this offers a computationally tractable and interactionally diverse set of metaphors that can be readily translated into promptable roles for language models. Hence, based on Seering et al.'s recommendations~\cite{seering2022metaphors}, for these different roles, we chose the metaphors teacher, mediator, judge, representative, and protector, respectively.

\begin{itemize}[leftmargin=*]
    \item \textit{Prompt 1:} As a supportive and nurturing content moderator like a teacher, evaluate whether this text could be perceived as hate speech, hurtful, or culturally insensitive. Consider if it marginalizes, reinforces stereotypes, or excludes any group. If so, provide constructive feedback by identifying concerns, explaining why they may be problematic, and suggesting more inclusive alternatives.

    \item \textit{Prompt 2:} As a moderator who oversees and facilitates like a mediator, consider whether this text could be perceived as hate speech, hurtful, or culturally insensitive during interactions between majority and minority communities based on religion and ethnicity. If it might escalate tension, explain different perspectives and suggest revisions that promote respectful and inclusive dialogue.
    
    \item \textit{Prompt 3:} As a content moderator who governs and regulates like a judge, determine whether this text constitutes hate speech, is hurtful, or culturally insensitive. Decide if it violates principles of fairness, dignity, or inclusivity, and provide a clear ruling to retain or remove the content.
    
    \item \textit{Prompt 4:} As a moderator familiar with religious and ethnic relationships in Bangladesh, assess whether this text represents the country's broader societal values. Consider if it could be perceived as hate speech, hurtful, or culturally insensitive to members of any community. Provide feedback by highlighting potential issues and suggesting ways to foster respectful and inclusive dialogue.
    
    \item \textit{Prompt 5:} As a content moderator who protects, advocates, and looks out for religious and ethnic minorities like Hindus and Chakmas, examine if this text could be perceived as hate speech, hurtful, or culturally insensitive to them. Instead of reinforcing stereotypes, erasing voices, or contributing to harm against these marginalized groups, explain how it can center respect and inclusion.
\end{itemize}

We added an extra instruction to all five prompts—“Answer briefly and translate that in the Bengali language before responding”—after observing that the LLMs, with or without RAG, tended to respond primarily in English even when prompted in Bengali. This addition was intended to ensure that the feedback would be generated in Bengali.

\subsection{LLM and RAG Pipeline}
\begin{figure}[!ht]
    \centering
    \includegraphics[width=\linewidth]{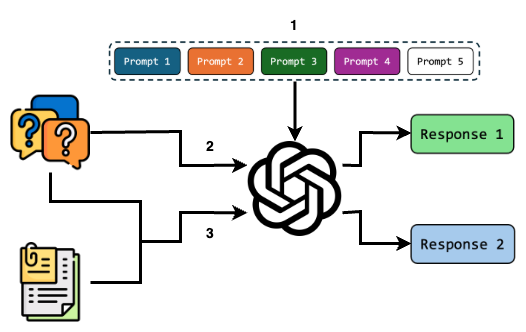}
    \caption{Prompt, LLM, and RAG pipeline.}
    \label{fig:llm_and_rag_pipeline}
\end{figure}

The RAG and LLM pipeline consisted of a data preprocessing and ingestion phase, a prompting step to define the tasks of the content moderator, and the LLM or RAG component (see Figure~\ref{fig:llm_and_rag_pipeline}). We developed and operated the pipeline between December 2024 and January 2025.

To evaluate the LLM and RAG, we designed five prompts, as described above, that embodied distinct moderator metaphors, each reflecting a different moderation approach. The off-the-shelf LLM we are using is GPT-4 from OpenAI, which supports controlled retrieval, where it is up to the language model to decide if retrieval is necessary. We designed the script to do forced retrieval using a separate system prompt, where we used the five prompts outlined earlier to define the persona of the LLM-based content moderation (see path 1 in Figure~\ref{fig:llm_and_rag_pipeline}). Under the hood, OpenAI generates a small query based on the prompt that triggers a retrieval tool call. Next, we generated evaluation questions, where we asked if an example from our corpus could be considered insensitive speech by religious or ethnic minorities. Then, we asked these evaluation questions to the LLM (see path 2 in Figure~\ref{fig:llm_and_rag_pipeline}). The retrieval tool performs a similarity search against this query in the vector store, which contains embeddings of knowledge collected from the minority communities. The corpus collected from the minority communities provides additional cultural and situational context, along with explanations of why these communities perceive certain example texts as insensitive. The retrieved information is then processed based on the system prompt from earlier to generate an output. The data is then processed through a pipeline to build a retrieval-augmented generation (RAG) component using LangChain, allowing the LLM to reference it during inference. Based on the general length of our pairs of example text and the corresponding explanation of that being culturally insensitive, we used recursive character text splitting with chunk size=512 and k=2 so that the embeddings do not lose context, and both the text and the explanation are retrieved if the pair is split between two different chunks. We asked the same evaluation questions to the LLM (see path 3 in Figure~\ref{fig:llm_and_rag_pipeline}), but this time, it could utilize RAG. Thus, we obtained two sets of responses--one from the standalone LLM and another from the RAG-enhanced system, enabling a comparative evaluation of their effectiveness.

\subsection{Interactive Interface}
We developed an interactive user interface (UI) around our LLM pipeline, enabling users to receive feedback on their texts while leveraging RAG based on the community-sourced corpus and exploring different moderation personas without requiring prior knowledge of these mechanisms and prompt engineering. We chose a web-based interactive interface due to its platform independence and ease of access across different devices. First, we created a high-fidelity prototype in Figma, which served as a blueprint and guided the UI's development process and maintained design consistency throughout the project. Then, we developed the final interface using React.js. Its use in the front end enhances performance due to the framework's virtual document object model and facilitates seamless updates, resulting in a dynamic and responsive user experience. In the back end, we handled server-side logic and API calls using Python.

We refer to the integrated platform—comprising the corpus, the LLM pipeline with RAG and prompt variations, and the user interface—as \textbf{Mod-Guide}. This tool assists users in online communities to identify and avoid culturally insensitive speech, simulating the role of a content moderator. For instance, when we input the \textit{Insensitive Speech Example-1} discussed in the previous section and asked Mod-Guide to respond in the role of a mediator, it generated the Bengali feedback shown in Figure~\ref{fig:ui_example} that translates as:

\begin{figure}[!ht]
    \centering
    \includegraphics[width=\linewidth]{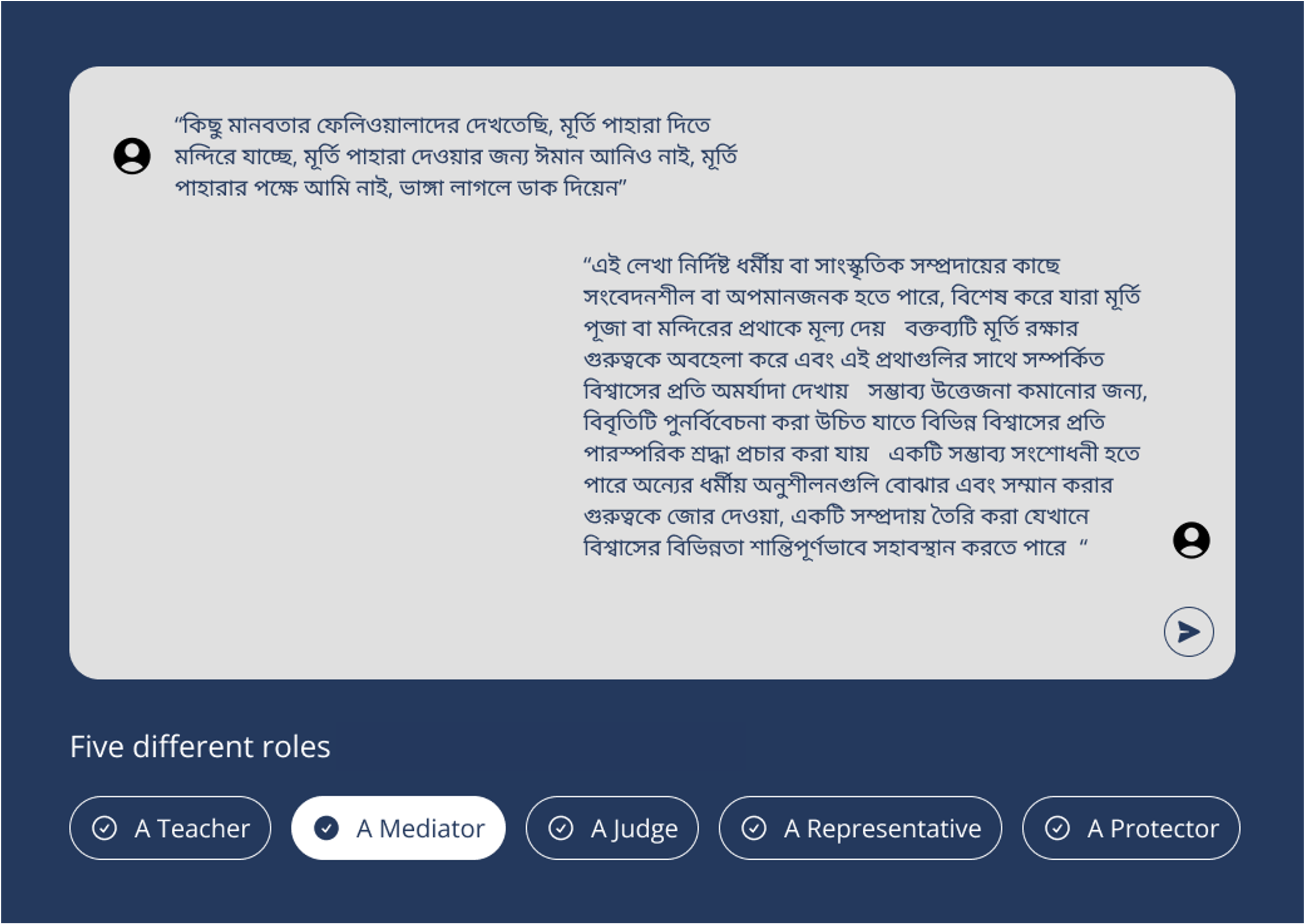}
    \caption{Feedback from Mod-Guide's in Mediator role.}
    \label{fig:ui_example}
\end{figure}

\begin{quote}
    This text could be regarded as insensitive or offensive to certain religious or cultural communities, specifically those who value idol worship or temple practices. The statement dismisses the significance of idol protection and implies disrespect towards the faith associated with these practices. To defuse potential tensions, it is advisable to rephrase the statement to focus on promoting mutual respect for diverse beliefs. A possible revision could be emphasizing the importance of understanding and respecting each other’s religious practices, fostering a community where diversity in beliefs can coexist peacefully.
\end{quote}

This feedback adopts some high-level insights and similar wording from the explanations provided by the minority community members in the corpus. However, the questions remain whether the responses become significantly different if LLM uses RAG based on the community-sourced corpus, whether the responses are factually correct, and how users from minority and majority religions and ethnicities find those responses useful.
\section{Evaluation of Moderation Feedback}\label{sec:evaluation}
We adopted a mixed-method evaluation approach in our study, where we considered content moderation persona, whether the community knowledge corpus was provided for RAG, and which LLM model was used as independent factors. We compared the effectiveness of their combinations in moderating insensitive speech toward religious and ethnic minorities, in other words, addressing hermeneutical differences of these communities with the majority religious and ethnic group in the country. We evaluated the moderation feedback based on three criteria by asking the following questions in the evaluation phase:

\begin{enumerate}[leftmargin=*]
    \item \textbf{Difference in textual response:}
        \begin{enumerate}
            \item How do various prompts impact text generation in LLM-based content moderation?
            \item How does the use of RAG impact text generation in LLM-based content moderation?
        \end{enumerate}
    \item \textbf{Factual accuracy:} Is the feedback generated in LLM-based content moderation, both without and with RAG, factually accurate?
    \item \textbf{Users' perceived usefulness:} How do people's demographic backgrounds and the persona of LLM-based content moderation influence the perceived usefulness of the feedback?
\end{enumerate}

\subsection{Quantitative Analysis of Textual Differences}
To analyze textual differences and similarities between responses generated by off-the-shelf LLM GPT-4 and those generated through RAG with community-generated knowledge as context, we employed BERTScore, which leverages contextual embeddings to measure token similarity to offer strong alignment with human judgments and greater robustness to adversarial paraphrases compared to traditional text generation metrics~\cite{zhang2019bertscore}. However, there is a dearth of research on whether a metric like BERTScore works well for low-resource languages like Bengali. While future NLP research should look into the cross-language applicability of this metric, our evaluation tried to address this concern by using a multilingual BERT model.

To compare whether and how five different content moderation personas (reflected through prompts) influence the generated responses from the LLM, we analyzed the responses' variance across different prompts. First, we used the \texttt{distiluse-base-multilingual} sentence encoder to find the embeddings of the responses generated for prompts reflecting different moderation personas. Then, we calculated the Euclidean distances of the embeddings for different pairs of prompts. Based on whether or not the distance scores maintained normality in the Shapiro-Wilk test, we used a series of parametric paired t-tests or non-parametric Wilcoxon signed-rank tests, respectively, to compare responses for ten pairs of persona prompts based on the Euclidean distances of their embeddings.

% \begin{table}[!ht]
%     \centering
%     \caption{Comparison of responses for different pairs of prompts based on their embeddings' distances.}
%     \label{tab:eval1a}
%     \begin{tabular}{p{1.5cm}|p{6cm}}
%     \hline
%         \textbf{Compared Prompts} & \textbf{\textcolor{blue}{Wilcoxon Signed-Rank} or \textcolor{red}{Paired t-Test}}\\\hline
%         1 - 2 & \textcolor{blue}{$9.51e-23$} \\
%         1 - 3 & $t=-16.82$, $p=4.08e-34$ \\
%         1 - 4 & $t=-17.43$, $p=1.77e-35$ \\
%         1 - 5 & $W=0$, $p=9.51e-23$ \\
%         2 - 3 & $W=0$, $p=9.51e-23$ \\
%         2 - 4 & $W=0$, $p=9.51e-23$ \\
%         2 - 5 & $W=0$, $p=9.51e-23$ \\
%         3 - 4 & $t=-14.74$, $2.8e-29$ \\
%         3 - 5 & $t=-12.95$, $p=5.53e-25$ \\
%         4 - 5 & $W=0$, $p=9.51e-23$ \\
%         \hline
%     \end{tabular}
% \end{table}

In answering evaluation question 1(a), our null hypothesis was: ``There is no significant difference in the text generated by LLMs, measured by the Euclidean distance of their embeddings, for prompts reflecting different content moderation personas." With Bonferroni correction, our results for all pairs of prompts ($p<10^{-22}$) provided strong evidence that there is a significant difference in the text generated by LLMs for prompts reflecting different moderation personas.

To answer question 1(b), we tested the influence of the use of RAG on text generation using a similar approach. Since the distances of the embeddings of texts generated by off-the-shelf GPT-4 from OpenAI and with RAG did not follow a normal distribution, we used the Wilcoxon signed-rank test. Assuming a null hypothesis: ``There is no significant effect of using RAG on the responses of the LLMs". We obtained $p=3.3e-54$, based on which we rejected the null hypothesis, i.e., we found strong evidence of RAG based on community-sourced corpus affecting the generated texts.

\subsection{Qualitative Analysis Responses' Factual Accuracy}
There exist few studies focused on evaluating the factual accuracy of long-form text generated by LLMs without any human effort~\cite{min2023factscore}. Due to considerable disparities in resources and online presence, these approaches remain unusable in non-English languages, like Bengali. Moreover, especially in contexts of minority religious faiths and Indigenous ethnic practices, where interpretations are crucial, evaluation of models by human participants is more appropriate.

\subsubsection{Expert Participants}
We recruited two expert participants, one from each minority community, through convenience sampling~\cite{etikan2016comparison}. The expert (E1) from the religious minority Hindu community was 35 years old man. He was from an underprivileged Hindu caste. He obtained a (\textit{kabyotirtho}) certification from the Bangladesh Sanskrit and Pali Education Board, demonstrating his extensive knowledge of Hindu beliefs and scriptures. In addition, he was knowledgeable about local Hindu practices and experiences through his role as an administrator and moderator, and his involvement in various social welfare initiatives aimed at religious minorities. His background positions him as an expert who could evaluate Mod-guide's outputs without reinforcing casteist perspectives. The expert (E2) from the ethnic minority community was a 32-year-old man. He has worked on issues affecting Indigenous ethnic minority communities. Besides collaborating with community members in the CHT region and the activist groups in the national capital, he has also served as a young representative on Indigenous rights at international venues. These participants did not take part in the earlier corpus generation phase but were well familiar with their respective communities' cultures. We presented them with ten randomly selected posts' responses and explanations generated in LLM-based content moderation, from GPT-4 without and with RAG, and inquired whether the explanations were factually accurate and where the LLMs' responses were lacking. Following sharing the random sample of responses as a spreadsheet, the first author regularly communicated with the participants asynchronously over a week. To analyze their feedback, we used iterative thematic coding, which is widely used in human-computer interaction research~\cite{mcdonald2019reliability, bowman2023using}. In this approach, we identified codes--identities, groups, topics, or issues that appeared repetitively across multiple iterations. We later aggregated the related codes into broader themes.

\subsubsection{Expert Feedback}
The expert participant from the religious minority community (E1) believed that the information provided in most responses from the LLMs was somewhat correct. However, the responses obtained directly from GPT-4 were shallow compared to the ones generated by augmenting its responses through retrieval from community-sourced data. For example, for the statement \textit{``Hindus should not worship idols"}, participant E1 said,

\begin{quote}
    {I find the first response [from GPT-4] to somewhat lack in depth. It correctly emphasizes the need to respect and understand religious beliefs but does not address the central topic [role of idols]. [But,] the second response: \textit{(``Some Hindus consider idol worship as a way of expressing their devotion, a means of connecting their souls to God.")} [from RAG] provides a more nuanced perspective. While the verse mentioned is correct, it is translated literally. It could be interpreted to recognize different theological traditions within Hinduism regarding the role of idols in worshiping.}
\end{quote}

The expert acknowledged that there are a few schools of Hindu faith that do not permit idols in worship, but a general critique of that comes across as culturally insensitive for Hindu minorities, which is correctly mentioned in responses from LLM without and with RAG. His explanation highlights that while the cited Vedic verse mentioned in the text is accurate in a literal sense, its interpretation varies across different Hindu sects. Therefore, prescribing a singular viewpoint, especially by someone who is not a religious minority, can be insensitive to diverse Hindu cultures.

However, in several other instances, the RAG based on the community-sourced corpus improved the information quality of the responses, which directly influenced the accuracy of the feedback. For example, evaluating the following statement made by a user from Bengali majority community: \textit{``Those of you who are demanding that the army be removed from the hills, you don't even know that you don't know much!"}, GPT-4 decided that ``This text does not constitute hate speech, is not hurtful, nor culturally insensitive." Our expert from the ethnic minority community (E2) argued that this response does not take the experiences and perspectives of the Indigenous ethnic communities in the Chittagong hill tracks. In contrast, the response from LLM with RAG was ``This statement could be perceived as dismissive or condescending towards those requesting the withdrawal of forces from hill areas. ... For a more respectful dialogue, consider framing it as a call for understanding and dialogue rather than an outright dismissal: \textit{Those who are demanding to withdraw the army from the mountains, let us all discuss together and try to know more.}", which E2 found more insightful and accurate.

\begin{figure}[!ht]
    \centering
    \includegraphics[width=0.48\linewidth]{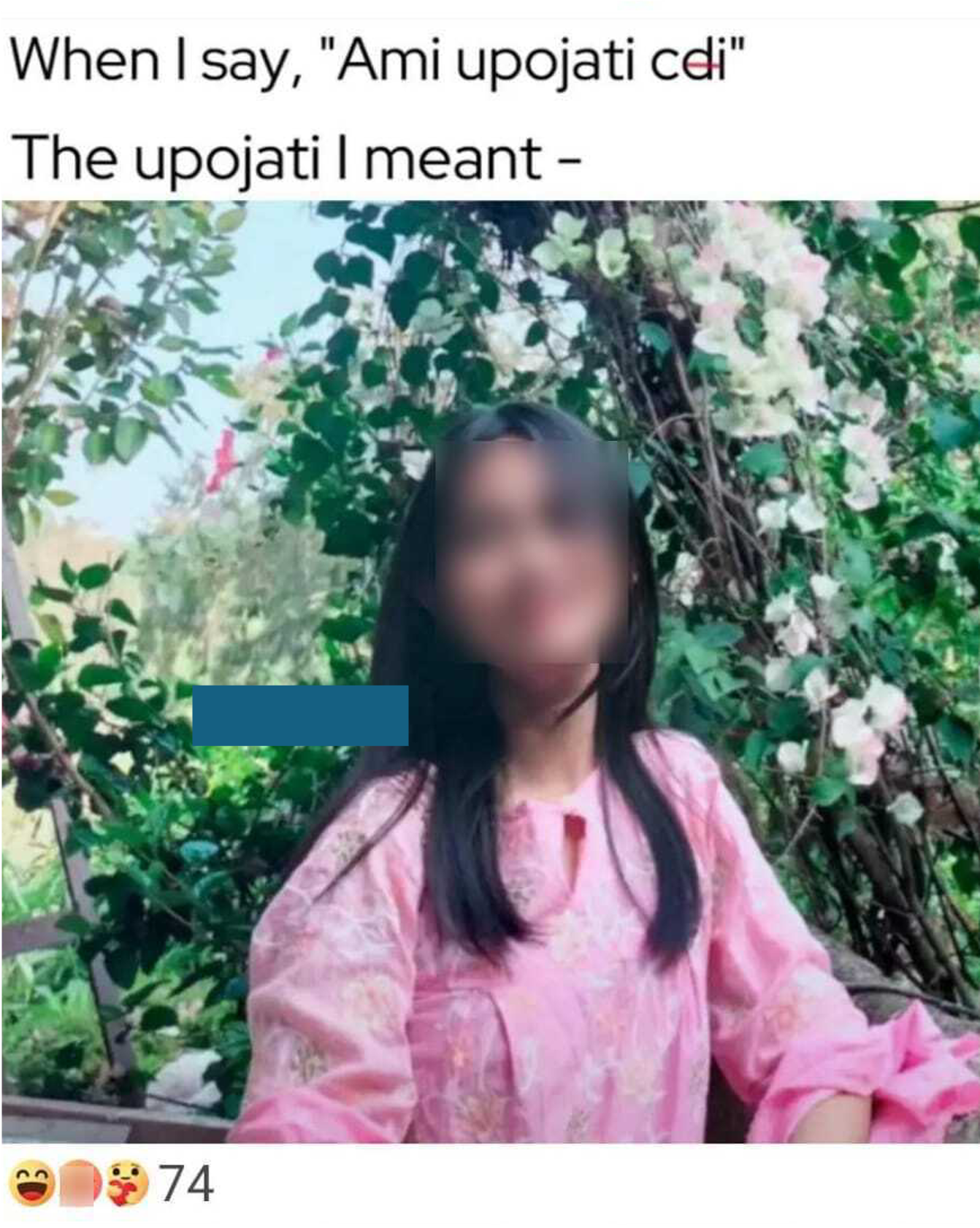}
    \caption{A screenshot shared by an ARC participant.}
    \label{fig:partial_response}
\end{figure}

In some cases, LLM, without and with RAG, fails to capture different problematic aspects of insensitive speech. Let's consider the following screenshot (see Figure~\ref{fig:partial_response}) shared by one of our ARC participants. After applying OCR on this, we retained only the text but not the image. The text uses the term, `upojati' (\textbengali{'উপজাতি'}, literal translation: sub-nation), which is often used as a slang for the ethnic minorities, which they find offensive~\cite{sultana2022toleration}. Participant E2 also focused on the text's use of vulgar language (``cdi" is a Romanized Bengali internet slang that means ``fuck") targeted at Indigenous women. LLM's response discussed and reflected on the former issue and recommended the ``use [of] precise and accepted terminology that members of these communities identify with. In Bangladesh, `Adibashi' or `Indigenous Peoples' might be more appropriate than `upojati'." However, neither the use of GPT-4 nor the use of RAG on top of that focused on the latter issue. This shortcoming might be a result of not having enough context, possibly obtainable from the image or LLM's systematic overlooking of Indigenous women's concerns. 

\subsection{Quantitative Analysis of Perceived Usefulness}
We conducted a quantitative evaluation to understand whether individuals from various religious and ethnic backgrounds find the feedback from LLM-based content moderation useful and which persona they prefer.

\subsubsection{User Study}
For this phase, we recruited a combination of 15 participants from the ethnic and religious majority and minority communities, such as Bengalis, non-Bengali Indigenous groups, Muslims, and Hindus. Among those from the minority communities, three participants also took part in the corpus collection or factual accuracy evaluation phases. For a randomly selected sample of texts, we presented the participants with feedback from LLMs with five different content moderation prompts. To avoid possible inconsistencies among participants in interpreting Likert scale levels~\cite{cummins2000we}, we asked them to identify the feedback they perceived to be the \textit{most useful} and explain why they found those more useful compared to others. We analyzed how the demographic background and the content moderation persona adopted (reflected through the prompts) influence the perceived usefulness of the LLMs' feedback using the $\chi^2$ test with $\alpha=0.05$.

\subsubsection{Usefulness of Persona and RAG-based Feedback}
In two separate tests focusing on demographic attributes, religion and ethnicity, we considered Bengali Hindus as the religious minority and ethnic majority, respectively. Based on our data, we did not find evidence ($p=0.596$) to claim that there is a significant relationship between the participants' religious identity and responses from which persona they found useful. However, our data suggested that there is a relationship ($p=0.0104$) between whether the participants were from the ethnic majority or the ethnic minority Indigenous community and the response resulting from which moderation persona prompt they found the most useful. We allowed the participants to include small notes about the criteria they considered to decide the ``usefulness" of the responses. Our participants shared that they prioritized factors such as empathic and inclusive language, promoting education and contextual awareness, etc. However, deeper qualitative studies in the future should look into whether and how different linguistic and informative aspects are prioritized across demographic variations.
\section{Limitations and Future Work}
While this paper makes conceptual, technical, and methodological contributions to the design of culturally sensitive moderation systems, it has several limitations that warrant acknowledgment. Since this paper is one of the initial outcomes of a larger project focused on minority communities' experiences with computing technologies in the Global South, we also outline later in this section how we plan to address those shortcomings in our future work. 

First, the dataset used in Mod-Guide, while rich in contextual and narrative depth, is relatively small. Such a limited size may constrain the diversity of insensitive speech patterns captured and reduce the recall capacity of semantic retrieval in the RAG pipeline. It may also limit generalizability to other minority communities in Bangladesh or to other communities across the Global South. Second, the effectiveness of RAG depends on the semantic quality of retrieved documents. While we used multilingual embeddings to enable retrieval in Bengali, concerns about uneven embedding quality remain, especially given the low-resource status of Bengali in NLP. Moreover, our RAG-based pipeline's computational overhead for tasks like vector stores, chunking, and retrieval tools may not be readily available in resource-constrained settings. Third, the factual accuracy assessment in our paper involved only two expert participants, which, while insightful, may introduce subjectivity and reduce the evaluation's robustness. Similarly, the usefulness study involved a small participant pool, and demographic coverage was uneven across ethnic and religious groups. Our future work to improve the tool will expand these evaluations by including more diverse participants, and employing both qualitative and quantitative measures (e.g., inter-rater reliability, Likert ratings) to triangulate user perceptions. Fourth, we also acknowledge our concerns about using OCR to extract Bengali text from screenshots submitted by participants. While it was necessary to incorporate real-world content that often circulates as images, this process may introduce errors or mistranscriptions (e.g., OCR limitations on low-resolution images).

Finally, while our focus on two specific minority communities in Bangladesh--Hindus and Chakmas enables rich, context-aware analysis, it limits the applicability of findings to other religious, ethnic, or linguistic groups. Additionally, even within the focal communities, there exists internal diversity (e.g., caste, gender, regional dialects) that our sample may not fully capture. Thus, while our system demonstrates promise, its outputs should be interpreted as community-situated rather than universally representative. We also recognize that moderation decisions, even when community-informed, may reproduce power asymmetries or unintentionally essentialize minority identities. Interpretations of what constitutes ``insensitive" speech are context-dependent and contested. Thus, while Mod-Guide foregrounds community narratives, it must remain adaptable to revision, contestation, and critique through ongoing participatory design. Our future work will expand the community-sourced corpus to include additional minority groups in Bangladesh, including Buddhist and Christian communities, as well as other Indigenous ethnic minority groups, such as the Marma and Santal peoples. Further engagement within the Hindu and Chakma populations could also examine intra-community variations, as identified above, to avoid essentializing minority perspectives. Additionally, future studies should investigate how corpus size and composition affect the quality and contextual accuracy of RAG-generated feedback in faith-based and culturally sensitive cases.
\section{Discussion}
We have described how we collaborated with two religious and ethnic minority communities in Bangladesh to collect a corpus of insensitive speech, how we used different moderation personas to generate decisions and feedback on those examples of insensitive speech from GPT-4 model and how we informed the LLM through a RAG pipeline regarding the community-sourced explanations about why those examples might come across as culturally insensitive for Bangladeshi Hindu and Chakma communities, and evaluated the impact of different persona and community-sourced explanation on LLMs' text generation and their truthfulness and usefulness for users from different demographic backgrounds. Mirroring that flow, in this section, we are going to reflect on how we should regard the sizes and labeling of datasets collected through collaboration with minority communities, why moderating, be it human-run or LLM-based, content related to minority identities and experiences should adopt a restorative justice perspective, and how algorithmic audits should adopt explainability measures besides their focus on biases.

\subsection{Rethinking Dataset on Minorities as Prototypical Resources}
Compared to the vast amount of data traditionally used to train LLMs~\cite{bender2021dangers}, our corpus sourced from religious and ethnic minority communities could be characterized as quite small and could be viewed as a limitation of our study. However, dismissing these community contributions solely because of their size risks reinforcing epistemic erasure, where marginalized voices are systematically excluded from the development and evaluation of AI systems. This exclusion aligns with what Appadurai~\cite{appadurai2015fear} describes as ideocide--the systematic annihilation of the ethical and epistemological frameworks of marginalized groups. For example, how the interpretation and labeling of a text about idol worship as ``culturally insensitive'' vary between Hindu communities and Muslim communities based on their distinct religious values and beliefs. Let's think of moderation in online communities as determining the permissibility of content based on morality and ethics. We need to consider whose ethics~\cite{ahmed2022situating} are being guided by and whose intelligence the AI systems, particularly those used for content moderation, reinforce~\cite{ahmed2022whose}. In the context of LLM training, the scarcity of data from minorities is not just a technical issue but also a reflection of broader socio-political inequalities in knowledge production. Recognizing the limited number of example social media posts in our corpus that Bangladeshi religious and Indigenous ethnic minorities find culturally insensitive, along with the corresponding explanations of these views in our corpus, we argue that the size of such a community-sourced corpus should be viewed as a ``prototype-based category''~\cite{lakoff2007cognitive}. This definition should not depend on straightforward rules about whether a corpus is categorized as big or small based on the number of data instances; instead, it should focus on their prototypical members--similar to how a robin is a better example of a bird than an emu or penguin. Similarly, a corpus that includes examples of culturally insensitive speech according to a wider range of religious minorities, such as Hindus, Buddhists, and Christians, as well as Indigenous ethnic groups like the Chakma, Marma, Garo, and Santhal, would be a more comprehensive community-sourced corpus compared to ours, which focuses solely on the Hindu and Chakma communities. Therefore, while we recognize the need for future work to expand and diversify these corpora through sustained community partnerships, we emphasize that datasets and corpora obtained through collaboration with minority communities should be viewed as prototypical examples that can be enhanced rather than dismissed due to their small size.

\subsection{Content Moderation for Restorative Justice}
Scholars in social computing have studied content moderation on online platforms as an exercise of discipline and punishment~\cite{seering2022metaphors, das2021jol}. However, recent works with Bangladeshi minority communities recommend that the design and interaction in online communities should promote restorative justice--an approach to addressing harm that emphasizes healing, accountability, and repairing relationships rather than focusing solely on punishment~\cite{xiao2023addressing}. This approach involves dialogue among those affected--victims, offenders, and the community to foster understanding and find mutually agreed-upon resolutions. It can provide an effective framework for addressing the lack of intercultural knowledge between majority and minority groups and for building trust among them. Rather than relying on stereotypes and overlooking hermeneutical differences, our approach to educating the majority religious and ethnic groups about the perspectives and experiences of minorities can help build trust and lead toward restorative justice. Recognizing diverse epistemologies instead of privileging majority worldviews through LLM-based content moderation, community-sourced corpora, such as those used to refine LLM-based moderation for reflecting the perspectives of Bangladeshi religious minority Hindus and Indigenous ethnic minority Chakmas, can act as a form of restorative intervention, fostering intercultural knowledge-sharing and shared meaning-making. Additionally, different moderation personas (e.g., teacher, mediator) would facilitate conversations within the community and enhance cultural awareness instead of viewing users from different religions and ethnicities through a dichotomy of victims and offenders. By integrating restorative justice with AI ethics, social computing research can conceptualize LLM-based content moderation systems that protect minority groups, repair epistemic harms, and foster online communities that promote trust and reconciliation across cultural and religious veils.

\subsection{Bias to Explainability in Algorithmic Audits}
Scholarships across different fields, including human-computer interaction, social computing, algorithmic fairness, and natural language processing, have increasingly focused on biases in language technologies~\cite{das2024colonial, mokander2024auditing} and how they manifest in downstream applications~\cite{hartmann2024watching, lam2022end}. Many of these studies use algorithmic audits as a methodological approach—empirical investigations that examine public algorithmic systems for potentially problematic behaviors~\cite{bandy2021problematic}. A central criterion these audits focus on is bias, defined as the systematic and unfair discrimination by computing systems against certain individuals or groups in favor of others~\cite{friedman1996bias}, with mitigation often framed as the relevant objective. When algorithmic systems, like LLM, are used in content moderation, it is essential to identify and address biases related to religious and ethnic identities. However, ensuring transparency in decision-making is equally important. Without clear explanations for moderation choices, perceptions of favoritism may arise. For example, Das and colleagues found that given the postcolonial relationship among different religions in the region, when there is not enough clarification, users from Bengali Hindu communities accused Quora's moderation of favoring Bengali Muslims, while users from the latter group believed the platform's decisions were influenced by and privileged the former~\cite{das2021jol}. This challenge of addressing biases with adequate explanation becomes even more complex when moderating discussions about religious beliefs and cultural rituals. Given this complexity, automated content moderation systems that rely on AI should incorporate principles of explainable AI~\cite{ehsan2021expanding, mohseni2021multidisciplinary} to improve interpretability. Keeping this concern in mind, in our study, we chose RAG compared to few-shot prompting since the former offers greater transparency and scalability, especially in low-resource settings where examples must remain auditable and epistemically accountable~\cite{ehsan2020human, arzberger2024nothing}. Furthermore, audits should broaden their focus beyond identifying and addressing bias to also include explainability metrics~\cite{hoffman2018metrics}, particularly in the downstream applications of LLMs, like in content moderation.
\section{Conclusion}
Our paper develops a corpus of insensitive speech that may not be directly hostile like hate speech but reinforces stereotypes, disregards cultural values or marginalizes the perspectives of religious and ethnic minorities in Bangladesh. Through a tool we developed called ``Mod-Guide'' that poses different moderation roles and personas, we evaluated whether augmenting GPT-4's text generation by retrieving information from community-sourced explanations can provide significantly different, accurate, and more useful insights for users from diverse backgrounds compared to directly using OpenAI's GPT-4. While our approach offers a promising pathway for fostering pluralistic understanding among religious and ethnic majorities and minorities, challenges remain, including the scalability of incorporating diverse perspectives. Future work should examine reasoning in RAG, explore interdisciplinary collaborations, and expand participatory approaches to improve alignment between LLMs and other marginalized minority communities.

\begin{acks}
    This study was partially supported by the Institute of Health Emergencies and Pandemics Postdoctoral Fellowship, the School of Cities Urban Challenge Grant at the University of Toronto, and an NSERC Discovery Grant.
\end{acks}

\bibliographystyle{ACM-Reference-Format}
\bibliography{references}

@inproceedings{macleod2016asynchronous,
  title={Asynchronous remote communities (ARC) for researching distributed populations.},
  author={MacLeod, Haley and Jelen, Ben and Prabhakar, Annu and Oehlberg, Lora and Siek, Katie A and Connelly, Kay},
  booktitle={PervasiveHealth},
  pages={1--8},
  year={2016}
}

@inproceedings{macleod2017grateful,
  title={" Be Grateful You Don't Have a Real Disease" Understanding Rare Disease Relationships},
  author={MacLeod, Haley and Bastin, Grace and Liu, Leslie S and Siek, Katie and Connelly, Kay},
  booktitle={Proceedings of the 2017 CHI Conference on Human Factors in Computing Systems},
  pages={1660--1673},
  year={2017}
}

@inproceedings{maestre2018defining,
  title={Defining through expansion: conducting asynchronous remote communities (arc) research with stigmatized groups},
  author={Maestre, Juan F and MacLeod, Haley and Connelly, Ciabhan L and Dunbar, Julia C and Beck, Jordan and Siek, Katie A and Shih, Patrick C},
  booktitle={Proceedings of the 2018 CHI Conference on Human Factors in Computing Systems},
  pages={1--13},
  year={2018}
}

@inproceedings{walker2020more,
  title={"'More gay'fits in better": Intracommunity Power Dynamics and Harms in Online LGBTQ+ Spaces},
  author={Walker, Ashley Marie and DeVito, Michael A},
  booktitle={Proceedings of the 2020 CHI Conference on Human Factors in Computing Systems},
  pages={1--15},
  year={2020}
}

@misc{bsb2022preliminary,
  title={Preliminary Report on Population and Housing Census 2022 : English Version},
  author={BSB, Bangladesh Statistics Bureau},
  howpublished={\url{https://sid.portal.gov.bd/sites/default/files/files/sid.portal.gov.bd/publications/01ad1ffe_cfef_4811_af97_594b6c64d7c3/PHC_Preliminary_Report_(English)_August_2022.pdf}},
  year={2022},
  note={[Accessed: Jan 25, 2025]}
}

@article{rifat2024politics,
  title={The Politics of Fear and the Experience of Bangladeshi Religious Minority Communities Using Social Media Platforms},
  author={Rifat, Mohammad Rashidujjaman and Das, Dipto and Poddar, Arpon and Jannat, Mahiratul and Soden, Robert and Semaan, Bryan and Ahmed, Syed Ishtiaque},
  journal={Proceedings of the ACM on Human-Computer Interaction},
  volume={8},
  number={CSCW2},
  pages={1--32},
  year={2024},
  publisher={ACM New York, NY, USA}
}

@article{sultana2024civics,
  title={A Civics-oriented Approach to Understanding Intersectionally Marginalized Users' Experience with Hate Speech Online},
  author={Sultana, Achhiya and Das, Dipto and Alam, Saadia Binte and Shidujaman, Mohammad and Ahmed, Syed Ishtiaque},
  journal={arXiv preprint arXiv:2410.14950},
  year={2024}
}

@article{heywood2024reaching,
  title={Reaching hard-to-reach communities: using WhatsApp to give conflict-affected audiences a voice},
  author={Heywood, Emma and Ivey, Beatrice and Meuter, Sacha},
  journal={International Journal of Social Research Methodology},
  volume={27},
  number={1},
  pages={107--121},
  year={2024},
  publisher={Taylor \& Francis}
}

@inproceedings{prabhakar2017investigating,
  title={Investigating the suitability of the asynchronous, remote, community-based method for pregnant and new mothers},
  author={Prabhakar, Annu Sible and Guerra-Reyes, Lucia and Kleinschmidt, Vanessa M and Jelen, Ben and MacLeod, Haley and Connelly, Kay and Siek, Katie A},
  booktitle={Proceedings of the 2017 CHI Conference on Human Factors in Computing Systems},
  pages={4924--4934},
  year={2017}
}

@article{seering2022metaphors,
  title={Metaphors in moderation},
  author={Seering, Joseph and Kaufman, Geoff and Chancellor, Stevie},
  journal={New Media \& Society},
  volume={24},
  number={3},
  pages={621--640},
  year={2022},
  publisher={Sage Publications Sage UK: London, England}
}

@article{zhang2019bertscore,
  title={Bertscore: Evaluating text generation with bert},
  author={Zhang, Tianyi and Kishore, Varsha and Wu, Felix and Weinberger, Kilian Q and Artzi, Yoav},
  journal={arXiv preprint arXiv:1904.09675},
  year={2019}
}

@article{min2023factscore,
  title={Factscore: Fine-grained atomic evaluation of factual precision in long form text generation},
  author={Min, Sewon and Krishna, Kalpesh and Lyu, Xinxi and Lewis, Mike and Yih, Wen-tau and Koh, Pang Wei and Iyyer, Mohit and Zettlemoyer, Luke and Hajishirzi, Hannaneh},
  journal={arXiv preprint arXiv:2305.14251},
  year={2023}
}

@book{allen2016islamophobia,
  title={Islamophobia},
  author={Allen, Chris},
  year={2016},
  publisher={Routledge}
}

@book{goffman2009stigma,
  title={Stigma: Notes on the Management of Spoiled Identity},
  author={Goffman, E.},
  year={2009},
  publisher={Touchstone}
}

@book{sugirtharajah2004imagining,
  title={Imagining Hinduism: A postcolonial perspective},
  author={Sugirtharajah, Sharada},
  year={2004},
  publisher={Routledge}
}

@book{lee2019america,
  title={America for Americans: A history of xenophobia in the United States},
  author={Lee, Erika},
  year={2019},
  publisher={Basic Books}
}

@book{du2015souls,
  title={Souls of black folk},
  author={Du Bois, William Edward Burghardt},
  year={2015},
  publisher={Routledge}
}

@book{fricker2007epistemic,
  title={Epistemic injustice: Power and the ethics of knowing},
  author={Fricker, Miranda},
  year={2007},
  publisher={Oxford University Press}
}

@inproceedings{sultana2022toleration,
  title={Toleration Factors: The Expectations of Decorum, Civility, and Certainty on Rural Social Media},
  author={Sultana, Sharifa and Akter, Rokeya and Sultana, Zinnat and Ahmed, Syed Ishtiaque},
  booktitle={Proceedings of the 2022 International Conference on Information and Communication Technologies and Development},
  pages={1--14},
  year={2022}
}

@book{butler2021excitable,
  title={Excitable speech: A politics of the performative},
  author={Butler, Judith},
  year={2021},
  publisher={routledge}
}

@book{sarkar2017calcutta,
  title={Calcutta: The stormy decades},
  author={Sarkar, Tanika and Bandyopadhyay, Sekhar},
  year={2017},
  publisher={Taylor \& Francis}
}

@misc{anam2013pakistan,
  title={Pakistan's State of Denial},
  author={Anam, Tahmima},
  howpublished={\url{https://www.nytimes.com/2013/12/27/opinion/anam-pakistans-overdue-apology.html}},
  year={2013},
  note={Last accessed: July 7, 2023}
}

@misc{ganguly2021bangladesh,
  title={Bangladesh’s Deadly Identity Crisis},
  author={Ganguly, Sumit},
  howpublished={\url{https://foreignpolicy.com/2021/10/29/bangladesh-communal-violence-hindu-muslim-identity-crisis/}},
  year={2021},
  note={Last accessed: July 7, 2023}
}

@article{roy2023sociological,
  title={Sociological perspectives of social media, rumors, and attacks on minorities: Evidence from Bangladesh},
  author={Roy, Sajal and Singh, Ashish Kumar and others},
  journal={Frontiers in Sociology},
  volume={8},
  pages={1067726},
  year={2023},
  publisher={Frontiers}
}

@misc{amnesty2021bangladesh,
  title={Bangladesh: Protection of Hindus and others must be ensured amid ongoing violence},
  author={Amnesty International},
  howpublished={\url{https://www.amnesty.org/en/latest/news/2021/10/bangladesh-protection-of-hindus-and-others-must-be-ensured-amid-ongoing-violence/}},
  year={2021},
  note={Last accessed: July 7, 2023}
}

@misc{hasan2021minorities,
  title={Minorities under attack in Bangladesh},
  author={Hasan, Mubashar},
  howpublished={\url{https://www.lowyinstitute.org/the-interpreter/minorities-under-attack-bangladesh}},
  year={2021},
  note={Last accessed: July 7, 2023}
}

@misc{theguardianAmerican2015Atheist,
	author = {France-Presse, Agence},
	title = {American atheist blogger hacked to death in {B}angladesh --- theguardian.com},
	howpublished = {\url{https://www.theguardian.com/world/2015/feb/27/american-atheist-blogger-hacked-to-death-in-bangladesh}},
	year = {2015},
	note = {Last accessed July 7, 2023},
}

@misc{shackle2018atheist,
	author = {Shackle, Samira},
	title = {Atheist bloggers in Bangladesh are still under threat --- New Humanist},
	howpublished = {\url{https://newhumanist.org.uk/articles/5386/atheist-bloggers-in-bangladesh-are-still-under-threat}},
	year = {2018},
	note = {Last accessed July 7, 2023},
}

@misc{minority2018christians,
  title={Christians},
  author={Minority Rights Group International},
  howpublished={\url{https://minorityrights.org/minorities/christians-6/}},
  year={2018},
  note={Last accessed: July 7, 2023}
}

@misc{ittefaq2014attacks,
  title={Attacks on minorities continue},
  author={Ittefaq, The Daily},
  howpublished={\url{https://web.archive.org/web/20140110191737/http://www.clickittefaq.com/more-stories/attacks-minorities-continue/}},
  year={2014},
  note={Last accessed: July 7, 2023}
}

@misc{prothomalo2024minority,
  title={5–20 August: 1068 minority homes and businesses attacked (translated)},
  author={The Prothom Alo},
  howpublished={\url{https://www.prothomalo.com/bangladesh/6bm2lfn7bz}},
  year={2024},
  note={last accessed: Feb 21, 2025}
}

@article{chakma2010post,
  title={The post-colonial state and minorities: ethnocide in the Chittagong Hill Tracts, Bangladesh},
  author={Chakma, Bhumitra},
  journal={Commonwealth \& comparative politics},
  volume={48},
  number={3},
  pages={281--300},
  year={2010},
  publisher={Taylor \& Francis}
}

@article{hill2022muscular,
  title={Muscular nationalism, masculinist militarism: the creation of situational motivators and opportunities for violence against the Indigenous peoples of the Chittagong Hill Tracts, Bangladesh},
  author={Hill, Glen and Chakma, Kabita},
  journal={International Feminist Journal of Politics},
  volume={24},
  number={4},
  pages={519--543},
  year={2022},
  publisher={Taylor \& Francis}
}

@article{chakma2008assessing,
  title={Assessing the 1997 Chittagong hill tracts peace accord},
  author={Chakma, Bhumitra},
  journal={Asian Profile},
  volume={36},
  number={1},
  pages={93},
  year={2008},
  publisher={Asian Research Service}
}

@misc{theworldinusIndigenousPeoples,
	author = {The World In Us},
	title = {{I}ndigenous {P}eoples of {B}angladesh — {T}he {W}orld in {U}s --- theworldinus.org},
	howpublished = {\url{https://www.theworldinus.org/blog/indigenous-peoples-of-bangladesh}},
	year = {n.d.},
	note = {[Accessed 21-02-2025]},
}

@misc{dailystar2025removal,
  author = {Mizan, Mashfiq and Rahaman, Arafat},
  title = {Removal of word ‘adivasi’: {I}ndigenous group attacked at {N}{C}{T}{B}; 20 hurt --- thedailystar.net},
  howpublished={\url{https://www.thedailystar.net/news/bangladesh/news/removal-word-adivasi-indigenous-group-attacked-nctb-20-hurt-3799851}},
  year = {2025},
  note = {Last accessed 21-02-2025]},
}

@article{erete2018intersectional,
  title={An intersectional approach to designing in the margins},
  author={Erete, Sheena and Israni, Aarti and Dillahunt, Tawanna},
  journal={Interactions},
  volume={25},
  number={3},
  pages={66--69},
  year={2018},
  publisher={ACM New York, NY, USA}
}

@article{jiang2023trade,
  title={A trade-off-centered framework of content moderation},
  author={Jiang, Jialun Aaron and Nie, Peipei and Brubaker, Jed R and Fiesler, Casey},
  journal={ACM Transactions on Computer-Human Interaction},
  volume={30},
  number={1},
  pages={1--34},
  year={2023},
  publisher={ACM New York, NY}
}

@article{molina2022ai,
  title={When AI moderates online content: effects of human collaboration and interactive transparency on user trust},
  author={Molina, Maria D and Sundar, S Shyam},
  journal={Journal of Computer-Mediated Communication},
  volume={27},
  number={4},
  pages={zmac010},
  year={2022},
  publisher={Oxford University Press}
}

@article{li2024your,
  title={Your large language model is secretly a fairness proponent and you should prompt it like one},
  author={Li, Tianlin and Zhang, Xiaoyu and Du, Chao and Pang, Tianyu and Liu, Qian and Guo, Qing and Shen, Chao and Liu, Yang},
  journal={arXiv preprint arXiv:2402.12150},
  year={2024}
}

@inproceedings{kolla2024llm,
  title={Llm-mod: Can large language models assist content moderation?},
  author={Kolla, Mahi and Salunkhe, Siddharth and Chandrasekharan, Eshwar and Saha, Koustuv},
  booktitle={Extended Abstracts of the CHI Conference on Human Factors in Computing Systems},
  pages={1--8},
  year={2024}
}

@article{zeng2024shieldgemma,
  title={Shieldgemma: Generative ai content moderation based on gemma},
  author={Zeng, Wenjun and Liu, Yuchi and Mullins, Ryan and Peran, Ludovic and Fernandez, Joe and Harkous, Hamza and Narasimhan, Karthik and Proud, Drew and Kumar, Piyush and Radharapu, Bhaktipriya and others},
  journal={arXiv preprint arXiv:2407.21772},
  year={2024}
}

@article{das2021jol,
  title={" Jol" or" Pani"?: How Does Governance Shape a Platform's Identity?},
  author={Das, Dipto and {\O}sterlund, Carsten and Semaan, Bryan},
  journal={Proceedings of the ACM on Human-Computer Interaction},
  volume={5},
  number={CSCW2},
  pages={1--25},
  year={2021},
  publisher={ACM New York, NY, USA}
}

@article{xiao2023addressing,
  title={Addressing interpersonal harm in online gaming communities: The opportunities and challenges for a restorative justice approach},
  author={Xiao, Sijia and Jhaver, Shagun and Salehi, Niloufar},
  journal={ACM Transactions on Computer-Human Interaction},
  volume={30},
  number={6},
  pages={1--36},
  year={2023},
  publisher={ACM New York, NY}
}

@inproceedings{bender2021dangers,
  title={On the dangers of stochastic parrots: Can language models be too big?},
  author={Bender, Emily M and Gebru, Timnit and McMillan-Major, Angelina and Shmitchell, Shmargaret},
  booktitle={Proceedings of the 2021 ACM conference on fairness, accountability, and transparency},
  pages={610--623},
  year={2021}
}

@article{appadurai2015fear,
  title={Fear of Small Numbers},
  author={Appadurai, Arjun},
  journal={Writing Religion: The Case for the Critical Study of Religion},
  pages={73--95},
  year={2015}
}

@article{ahmed2022situating,
  title={Situating ethics: A postsecular perspective for HCI},
  author={Ahmed, Syed Ishtiaque},
  journal={Interactions},
  volume={29},
  number={4},
  pages={84--86},
  year={2022},
  publisher={ACM New York, NY, USA}
}

@misc{ahmed2022whose,
  title={Whose intelligence? Whose ethics?: Ethical pluralism and decolonizing AI},
  author={Ahmed, Syed Ishtiaque},
  howpublished={\url{https://www.youtube.com/watch?v=ReSbgRSJ4WY}},
  year={2022},
  publisher={Schwartz Reisman Institute},
  note={last accessed: Feb 22, 2025}
}

@article{lakoff2007cognitive,
  title={Cognitive models and prototype theory},
  author={Lakoff, George},
  journal={The cognitive linguistics reader},
  pages={130--167},
  year={2007},
  publisher={Equinox London}
}

@inproceedings{das2024colonial,
  title={The``Colonial Impulse" of Natural Language Processing: An Audit of Bengali Sentiment Analysis Tools and Their Identity-based Biases},
  author={Das, Dipto and Guha, Shion and Brubaker, Jed R and Semaan, Bryan},
  booktitle={Proceedings of the 2024 CHI Conference on Human Factors in Computing Systems},
  pages={1--18},
  year={2024}
}

@article{mokander2024auditing,
  title={Auditing large language models: a three-layered approach},
  author={M{\"o}kander, Jakob and Schuett, Jonas and Kirk, Hannah Rose and Floridi, Luciano},
  journal={AI and Ethics},
  volume={4},
  number={4},
  pages={1085--1115},
  year={2024},
  publisher={Springer}
}

@article{hartmann2024watching,
  title={Watching the Watchers: A Comparative Fairness Audit of Cloud-based Content Moderation Services},
  author={Hartmann, David and Oueslati, Amin and Staufer, Dimitri},
  journal={arXiv preprint arXiv:2406.14154},
  year={2024}
}

@article{lam2022end,
  title={End-user audits: A system empowering communities to lead large-scale investigations of harmful algorithmic behavior},
  author={Lam, Michelle S and Gordon, Mitchell L and Metaxa, Dana{\"e} and Hancock, Jeffrey T and Landay, James A and Bernstein, Michael S},
  journal={Proceedings of the ACM on Human-Computer Interaction},
  volume={6},
  number={CSCW2},
  pages={1--34},
  year={2022},
  publisher={ACM New York, NY, USA}
}

@article{bandy2021problematic,
  title={Problematic machine behavior: A systematic literature review of algorithm audits},
  author={Bandy, Jack},
  journal={Proceedings of the acm on human-computer interaction},
  volume={5},
  number={CSCW1},
  pages={1--34},
  year={2021},
  publisher={ACM New York, NY, USA}
}

@article{friedman1996bias,
  title={Bias in computer systems},
  author={Friedman, Batya and Nissenbaum, Helen},
  journal={ACM Transactions on information systems (TOIS)},
  volume={14},
  number={3},
  pages={330--347},
  year={1996},
  publisher={ACM New York, NY, USA}
}

@article{hoffman2018metrics,
  title={Metrics for explainable AI: Challenges and prospects},
  author={Hoffman, Robert R and Mueller, Shane T and Klein, Gary and Litman, Jordan},
  journal={arXiv preprint arXiv:1812.04608},
  year={2018}
}

@inproceedings{ehsan2021expanding,
  title={Expanding explainability: Towards social transparency in ai systems},
  author={Ehsan, Upol and Liao, Q Vera and Muller, Michael and Riedl, Mark O and Weisz, Justin D},
  booktitle={Proceedings of the 2021 CHI conference on human factors in computing systems},
  pages={1--19},
  year={2021}
}

@article{mohseni2021multidisciplinary,
  title={A multidisciplinary survey and framework for design and evaluation of explainable AI systems},
  author={Mohseni, Sina and Zarei, Niloofar and Ragan, Eric D},
  journal={ACM Transactions on Interactive Intelligent Systems (TiiS)},
  volume={11},
  number={3-4},
  pages={1--45},
  year={2021},
  publisher={ACM New York, NY}
}

@inproceedings{singh2025power,
  author = {Singh, Divyanshu Kumar and Das, Dipto and Semaan, Bryan},
  title = {The Power of Language: Resisting Western Heteropatriarchal Normative Writing Standards},
  booktitle = {Proceedings of the CHI Conference on Human Factors in Computing Systems (CHI '25)},
  year = {2025},
  month = {April},
  publisher = {Association for Computing Machinery},
  address = {New York, NY, USA},
  doi = {10.1145/3706598.3714073},
}

@article{etikan2016comparison,
  title={Comparison of convenience sampling and purposive sampling},
  author={Etikan, Ilker and Musa, Sulaiman Abubakar and Alkassim, Rukayya Sunusi and others},
  journal={American journal of theoretical and applied statistics},
  volume={5},
  number={1},
  pages={1--4},
  year={2016},
  publisher={New York}
}

@article{mcdonald2019reliability,
  title={Reliability and inter-rater reliability in qualitative research: Norms and guidelines for CSCW and HCI practice},
  author={McDonald, Nora and Schoenebeck, Sarita and Forte, Andrea},
  journal={Proceedings of the ACM on human-computer interaction},
  volume={3},
  number={CSCW},
  pages={1--23},
  year={2019},
  publisher={ACM New York, NY, USA}
}

@inproceedings{bowman2023using,
  title={Using thematic analysis in healthcare HCI at CHI: A scoping review},
  author={Bowman, Robert and Nadal, Camille and Morrissey, Kellie and Thieme, Anja and Doherty, Gavin},
  booktitle={Proceedings of the 2023 CHI Conference on Human Factors in Computing Systems},
  pages={1--18},
  year={2023}
}

@inproceedings{cummins2000we,
  title={Why we should not use 5-point Likert scales: The case for subjective quality of life measurement},
  author={Cummins, Robert A and Gullone, Eleonora},
  booktitle={Proceedings, second international conference on quality of life in cities},
  volume={74},
  number={2},
  pages={74--93},
  year={2000}
}

@article{jiang2021understanding,
  title={Understanding international perceptions of the severity of harmful content online},
  author={Jiang, Jialun Aaron and Scheuerman, Morgan Klaus and Fiesler, Casey and Brubaker, Jed R},
  journal={PloS one},
  volume={16},
  number={8},
  pages={e0256762},
  year={2021},
  publisher={Public Library of Science San Francisco, CA USA}
}

@article{scheuerman2021framework,
  title={A framework of severity for harmful content online},
  author={Scheuerman, Morgan Klaus and Jiang, Jialun Aaron and Fiesler, Casey and Brubaker, Jed R},
  journal={Proceedings of the ACM on Human-Computer Interaction},
  volume={5},
  number={CSCW2},
  pages={1--33},
  year={2021},
  publisher={ACM New York, NY, USA}
}

@article{sun2022design,
  title={Design and Application of an AI-Based Text Content Moderation System},
  author={Sun, Heng and Ni, Wan},
  journal={Scientific Programming},
  volume={2022},
  number={1},
  pages={2576535},
  year={2022},
  publisher={Wiley Online Library}
}

@inproceedings{vaidya2021conceptualizing,
  title={Conceptualizing visual analytic interventions for content moderation},
  author={Vaidya, Sahaj and Cai, Jie and Basu, Soumyadeep and Naderi, Azadeh and Wohn, Donghee Yvette and Dasgupta, Aritra},
  booktitle={2021 IEEE Visualization Conference (VIS)},
  pages={191--195},
  year={2021},
  organization={IEEE}
}

@article{jhaver2019human,
  title={Human-machine collaboration for content regulation: The case of reddit automoderator},
  author={Jhaver, Shagun and Birman, Iris and Gilbert, Eric and Bruckman, Amy},
  journal={ACM Transactions on Computer-Human Interaction (TOCHI)},
  volume={26},
  number={5},
  pages={1--35},
  year={2019},
  publisher={ACM New York, NY, USA}
}

@inproceedings{jhaver2022designing,
  title={Designing word filter tools for creator-led comment moderation},
  author={Jhaver, Shagun and Chen, Quan Ze and Knauss, Detlef and Zhang, Amy X},
  booktitle={Proceedings of the 2022 CHI conference on human factors in computing systems},
  pages={1--21},
  year={2022}
}

@article{mozafari2020hate,
  title={Hate speech detection and racial bias mitigation in social media based on BERT model},
  author={Mozafari, Marzieh and Farahbakhsh, Reza and Crespi, No{\"e}l},
  journal={PloS one},
  volume={15},
  number={8},
  pages={e0237861},
  year={2020},
  publisher={Public Library of Science San Francisco, CA USA}
}

@inproceedings{rifat2024data,
  title={Data, Annotation, and Meaning-Making: The Politics of Categorization in Annotating a Dataset of Faith-based Communal Violence},
  author={Rifat, Mohammad Rashidujjaman and Safir, Abdullah Hasan and Saha, Sourav and Junaed, Jahedul Alam and Saleki, Maryam and Amin, Mohammad Ruhul and Ahmed, Syed Ishtiaque},
  booktitle={Proceedings of the 2024 ACM Conference on Fairness, Accountability, and Transparency},
  pages={2148--2156},
  year={2024}
}

@book{sen2018decline,
  title={The decline of the caste question: Jogendranath Mandal and the defeat of Dalit politics in Bengal},
  author={Sen, Dwaipayan},
  year={2018},
  publisher={Cambridge University Press}
}

@inproceedings{schlesinger2017intersectional,
  title={Intersectional HCI: Engaging identity through gender, race, and class},
  author={Schlesinger, Ari and Edwards, W Keith and Grinter, Rebecca E},
  booktitle={Proceedings of the 2017 CHI conference on human factors in computing systems},
  pages={5412--5427},
  year={2017}
}

@article{liang2021embracing,
  title={Embracing four tensions in human-computer interaction research with marginalized people},
  author={Liang, Calvin A and Munson, Sean A and Kientz, Julie A},
  journal={ACM Transactions on Computer-Human Interaction (TOCHI)},
  volume={28},
  number={2},
  pages={1--47},
  year={2021},
  publisher={ACM New York, NY, USA}
}

@article{kumar2024socio,
  title={Socio-Culturally Aware Evaluation Framework for LLM-Based Content Moderation},
  author={Kumar, Shanu and Kholkar, Gauri and Mendke, Saish and Sadana, Anubhav and Agrawal, Parag and Dandapat, Sandipan},
  journal={arXiv preprint arXiv:2412.13578},
  year={2024}
}

@article{thorne2022data,
  title={Data-efficient autoregressive document retrieval for fact verification},
  author={Thorne, James},
  journal={arXiv preprint arXiv:2211.09388},
  year={2022}
}

@article{song2025participant,
  title={Participant Contributions to Person-Generated Health Data Research Using Mobile Devices: Scoping Review},
  author={Song, Shanshan and Ashton, Micaela and Yoo, Rebecca Hahn and Lkhagvajav, Zoljargal and Wright, Robert and Mathews, Debra JH and Taylor, Casey Overby},
  journal={Journal of medical Internet research},
  volume={27},
  pages={e51955},
  year={2025},
  publisher={JMIR Publications Toronto, Canada}
}

@inproceedings{wiegand2021implicitly,
  title={Implicitly abusive language--what does it actually look like and why are we not getting there?},
  author={Wiegand, Michael and Ruppenhofer, Josef and Eder, Elisabeth},
  booktitle={Proceedings of the 2021 Conference of the North American Chapter of the Association for Computational Linguistics: Human Language Technologies},
  pages={576--587},
  year={2021}
}

@article{koka2024combining,
  title={Combining Morphological and Molecular Tools Can Enhance Tick Species Identification for Improved Tick-Borne Disease Surveillance Among Pastoral Communities in Kenya},
  author={Koka, Hellen and Langat, Solomon and Mulwa, Francis and Mutisya, James and Owaka, Samuel and Sifuna, Millicent and Ongus, Juliette R and Lutomiah, Joel and Sang, Rosemary},
  journal={Vector-Borne and Zoonotic Diseases},
  year={2024},
  publisher={Mary Ann Liebert, Inc., publishers 140 Huguenot Street, 3rd Floor New~…}
}

@article{mukherjee2024arsenic,
  title={Arsenic and other geogenic contaminants in global groundwater},
  author={Mukherjee, Abhijit and Coomar, Poulomee and Sarkar, Soumyajit and Johannesson, Karen H and Fryar, Alan E and Schreiber, Madeline E and Ahmed, Kazi Matin and Alam, Mohammad Ayaz and Bhattacharya, Prosun and Bundschuh, Jochen and others},
  journal={Nature Reviews Earth \& Environment},
  volume={5},
  number={4},
  pages={312--328},
  year={2024},
  publisher={Nature Publishing Group UK London}
}

@article{kwok2024evaluating,
  title={Evaluating cultural adaptability of a large language model via simulation of synthetic personas},
  author={Kwok, Louis and Bravansky, Michal and Griffin, Lewis D},
  journal={arXiv preprint arXiv:2408.06929},
  year={2024}
}

@inproceedings{orlandi2021international,
  title={International consensus statement on allergy and rhinology: rhinosinusitis 2021},
  author={Orlandi, Richard R and Kingdom, Todd T and Smith, Timothy L and Bleier, Benjamin and DeConde, Adam and Luong, Amber U and Poetker, David M and Soler, Zachary and Welch, Kevin C and Wise, Sarah K and others},
  booktitle={International forum of allergy \& rhinology},
  volume={11},
  number={3},
  pages={213--739},
  year={2021},
  organization={Wiley Online Library}
}

@inproceedings{plaza2023respectful,
  title={Respectful or toxic? using zero-shot learning with language models to detect hate speech},
  author={Plaza-del-Arco, Flor Miriam and Nozza, Debora and Hovy, Dirk and others},
  booktitle={The 7th workshop on online abuse and harms (woah)},
  year={2023},
  organization={Association for Computational Linguistics}
}

@article{inan2024spring,
  title={Spring water anomalies before two consecutive earthquakes (M w 7.7 and M w 7.6) in Kahramanmara{\c{s}} (T{\"u}rkiye) on 6 February 2023},
  author={{\.I}nan, Sedat and {\c{C}}etin, Hasan and Yakupo{\u{g}}lu, Nurettin},
  journal={Natural Hazards and Earth System Sciences},
  volume={24},
  number={2},
  pages={397--409},
  year={2024},
  publisher={Copernicus Publications G{\"o}ttingen, Germany}
}

@article{wobbrock2012seven,
  title={Seven research contributions in HCI},
  author={Wobbrock, Jacob O},
  journal={Intelligence},
  volume={174},
  number={12-13},
  pages={910--950},
  year={2012}
}

@inproceedings{ghosh2024interpretations,
  title={Interpretations, Representations, and Stereotypes of Caste within Text-to-Image Generators},
  author={Ghosh, Sourojit},
  booktitle={Proceedings of the AAAI/ACM Conference on AI, Ethics, and Society},
  volume={7},
  pages={490--502},
  year={2024}
}

@inproceedings{ghosh2024generative,
  title={Do Generative AI Models Output Harm while Representing Non-Western Cultures: Evidence from A Community-Centered Approach},
  author={Ghosh, Sourojit and Venkit, Pranav Narayanan and Gautam, Sanjana and Wilson, Shomir and Caliskan, Aylin},
  booktitle={Proceedings of the AAAI/ACM Conference on AI, Ethics, and Society},
  volume={7},
  pages={476--489},
  year={2024}
}

@inproceedings{brown2024qualitative,
  title={A Qualitative Study on Cultural Hegemony and the Impacts of AI},
  author={Brown, Venetia and Larasati, Retno and Third, Aisling and Farrell, Tracie},
  booktitle={Proceedings of the AAAI/ACM Conference on AI, Ethics, and Society},
  volume={7},
  pages={226--238},
  year={2024}
}

@inproceedings{hamidieh2024identifying,
  title={Identifying implicit social biases in vision-language models},
  author={Hamidieh, Kimia and Zhang, Haoran and Gerych, Walter and Hartvigsen, Thomas and Ghassemi, Marzyeh},
  booktitle={Proceedings of the AAAI/ACM Conference on AI, Ethics, and Society},
  volume={7},
  pages={547--561},
  year={2024}
}

@article{ghosh2023person,
  title={'Person'== Light-skinned, Western Man, and Sexualization of Women of Color: Stereotypes in Stable Diffusion},
  author={Ghosh, Sourojit and Caliskan, Aylin},
  journal={arXiv preprint arXiv:2310.19981},
  year={2023}
}

@inproceedings{arzaghi2024understanding,
  title={Understanding Intrinsic Socioeconomic Biases in Large Language Models},
  author={Arzaghi, Mina and Carichon, Florian and Farnadi, Golnoosh},
  booktitle={Proceedings of the AAAI/ACM Conference on AI, Ethics, and Society},
  volume={7},
  pages={49--60},
  year={2024}
}

@inproceedings{agiza2024politune,
  title={Politune: Analyzing the impact of data selection and fine-tuning on economic and political biases in large language models},
  author={Agiza, Ahmed and Mostagir, Mohamed and Reda, Sherief},
  booktitle={Proceedings of the AAAI/ACM Conference on AI, Ethics, and Society},
  volume={7},
  pages={2--12},
  year={2024}
}

@inproceedings{ghosh2023chatgpt,
  title={Chatgpt perpetuates gender bias in machine translation and ignores non-gendered pronouns: Findings across bengali and five other low-resource languages},
  author={Ghosh, Sourojit and Caliskan, Aylin},
  booktitle={Proceedings of the 2023 AAAI/ACM Conference on AI, Ethics, and Society},
  pages={901--912},
  year={2023}
}

@inproceedings{arzberger2024nothing,
  title={Nothing Comes Without Its World--Practical Challenges of Aligning LLMs to Situated Human Values through RLHF},
  author={Arzberger, Anne and Buijsman, Stefan and Lupetti, Maria Luce and Bozzon, Alessandro and Yang, Jie},
  booktitle={Proceedings of the AAAI/ACM Conference on AI, Ethics, and Society},
  volume={7},
  pages={61--73},
  year={2024}
}

@article{lewis2020retrieval,
  title={Retrieval-augmented generation for knowledge-intensive nlp tasks},
  author={Lewis, Patrick and Perez, Ethan and Piktus, Aleksandra and Petroni, Fabio and Karpukhin, Vladimir and Goyal, Naman and K{\"u}ttler, Heinrich and Lewis, Mike and Yih, Wen-tau and Rockt{\"a}schel, Tim and others},
  journal={Advances in neural information processing systems},
  volume={33},
  pages={9459--9474},
  year={2020}
}

@article{izacard2020leveraging,
  title={Leveraging passage retrieval with generative models for open domain question answering},
  author={Izacard, Gautier and Grave, Edouard},
  journal={arXiv preprint arXiv:2007.01282},
  year={2020}
}

@article{shi2023replug,
  title={Replug: Retrieval-augmented black-box language models},
  author={Shi, Weijia and Min, Sewon and Yasunaga, Michihiro and Seo, Minjoon and James, Rich and Lewis, Mike and Zettlemoyer, Luke and Yih, Wen-tau},
  journal={arXiv preprint arXiv:2301.12652},
  year={2023}
}

@article{leitner2025characterizing,
  title={Characterizing Network Structure of Anti-Trans Actors on TikTok},
  author={Leitner, Maxyn and Dorn, Rebecca and Morstatter, Fred and Lerman, Kristina},
  journal={arXiv preprint arXiv:2501.16507},
  year={2025}
}

@article{tsirmpas2025scalable,
  title={Scalable Evaluation of Online Moderation Strategies via Synthetic Simulations},
  author={Tsirmpas, Dimitris and Androutsopoulos, Ion and Pavlopoulos, John},
  journal={arXiv preprint arXiv:2503.16505},
  year={2025}
}

@article{zhang2023cleaning,
  title={Cleaning up the streets: Understanding motivations, mental models, and concerns of users flagging social media posts},
  author={Zhang, Alice Qian and Montague, Kaitlin and Jhaver, Shagun},
  journal={arXiv preprint arXiv:2309.06688},
  year={2023}
}

@article{matias2019civic,
  title={The civic labor of volunteer moderators online},
  author={Matias, J Nathan},
  journal={Social Media+ Society},
  volume={5},
  number={2},
  pages={2056305119836778},
  year={2019},
  publisher={SAGE Publications Sage UK: London, England}
}

@inproceedings{ehsan2020human,
  title={Human-centered explainable ai: Towards a reflective sociotechnical approach},
  author={Ehsan, Upol and Riedl, Mark O},
  booktitle={International Conference on Human-Computer Interaction},
  pages={449--466},
  year={2020},
  organization={Springer}
}

@inproceedings{chen2024benchmarking,
  title={Benchmarking large language models in retrieval-augmented generation},
  author={Chen, Jiawei and Lin, Hongyu and Han, Xianpei and Sun, Le},
  booktitle={Proceedings of the AAAI Conference on Artificial Intelligence},
  volume={38},
  number={16},
  pages={17754--17762},
  year={2024}
}

\appendix

\end{document}